\documentclass[12pt]{mn2e}
\usepackage{graphicx}
\usepackage{epsf}
\usepackage{lscape}
\usepackage{longtable}
\usepackage{rotating}
\usepackage{color}

\newcommand{\aap}{A\&A}

\newcommand{\apj}{ApJ}
\newcommand{\apjl}{ApJ}

\newcommand{\mnras}{MNRAS}
\newcommand{\aj}{AJ}

\newcommand{\nat}{Nat}

\begin{document}
\topmargin -0.5in 

\title[The UV Properties of Star Forming Galaxies I]{The UV Properties of Star Forming Galaxies I: {\em HST} WFC3 Observations of Very-high Redshift Galaxies}

\author[Stephen M. Wilkins, et al.\ ]  
{
Stephen M. Wilkins$^{1}$\thanks{E-mail: stephen.wilkins@astro.ox.ac.uk}, Andrew J. Bunker$^{1}$, Elizabeth Stanway$^{2,3}$, Silvio Lorenzoni$^{1}$, \newauthor Joseph Caruana$^{1}$ \\
$^1$\,Department of Physics, University of Oxford, Denys Wilkinson Building, Keble Road, OX1 3RH, U.K. \\
$^2$\,Department of Physics, University of Warwick, Gibbet Hill Road, Coventry, CV4 7AL, U.K.\\
$^3$\,H H Wills Physics Laboratory, Tyndall Avenue, Bristol, BS8 1TL, U.K.\\}
\maketitle

\begin{abstract}
The acquisition of deep Near-IR imaging with Wide Field Camera 3 on the Hubble Space Telescope has provided the opportunity to study the very-high redshift Universe. For galaxies up to $z\approx 7.7$ sufficient wavelength coverage exists to probe the rest-frame ultraviolet (UV) continuum without contamination from either Lyman-$\alpha$ emission or the Lyman-$\alpha$ break. In this work we use Near-IR imaging to measure the rest-frame UV continuum colours of galaxies at $4.7<z<7.7$. We are have carefully defined a colour-colour selection to minimise any inherent bias in the measured UV continuum slope for the drop-out samples. For the highest-redshift sample ($6.7<z<7.7$), selected as $z_{f850lp}$-band dropouts, we find mean UV continuum colours approximately equal to zero (AB), consistent with a dust-free, solar metallicity, star forming population (or a moderately dusty population of low metallicity). At lower-redshift we find that the mean UV continuum colours of galaxies (over the same luminosity range) are redder, and that galaxies with higher luminosities are also slightly redder on average. One interpretation of this is that lower-redshift and more luminous galaxies are dustier, however this interpretation is complicated by the effects of the star formation history and metallicity and potentially the initial mass function on the UV continuum colours.
\end{abstract} 


\section{Introduction}

The addition of Wide Field Camera 3 (WFC3) to the Hubble Space Telescope ({\em HST}) provides the sensitivity to efficiently probe the Universe at observed near-Infrared (NIR) wavelengths. These abilities are particularly important for studies of the very-high redshift Universe, as the rest-frame UV and optical emission is redshifted into the NIR. 

Broadband NIR imaging such as that obtained by WFC3 can be exploited by taking advantage of the redshifted Lyman-$\alpha$ break (e.g. at $z=7.2$ Lyman-$\alpha_{\rm obs}\approx 1\mu m$) to allow the identification of numerous high-redshift ($6<z<9$) star forming galaxy candidates (e.g. Oesch et al. 2010, Bouwens et al. 2010b, Bunker et al. 2010, McLure et al. 2010, Wilkins et al. 2010, Finkelstein et al. 2010, Wilkins et al. 2011, Bouwens et al. 2010c, Lorenzoni et al. 2011, Mclure et al. 2011). From these observations it is possible to constrain the rest-frame ultraviolet (UV) luminosity function, and thus the UV luminosity density. Making assumptions regarding the composition and properties of the stellar populations (i.e. metallicity, star formation history, initial mass function) and dust attenuation we can constrain both the star formation rate and ionising photon luminosity density (e.g. Wilkins et al. 2011).

The deep WFC3 NIR imaging also allows the measurement of the rest-frame UV continuum colours of high-redshift $4.7<z<7.7$ objects without the possibility of contamination from either the Lyman-$\alpha$ break or Lyman-$\alpha$ emission. The UV continuum colour is theoretically predicted and empirically confirmed to be sensitive to the amount of dust reddening (e.g. Meurer et al. 1999, Pannella et al. 2009), potentially making it a useful diagnostic of the dust attenuation and thus providing a method for constraining {\em intrinsic} star formation rate density (as opposed to the {\em observed} star formation density). Indeed several recent studies have investigated the UV continuum slopes of high-redshift galaxies (e.g. Stanway, McMahon \& Bunker 2005, Bouwens et al. 2009, Bunker et al. 2010, Bouwens et al. 2010a, Dunlop et al. 2011) typically finding blue UV continuum slopes consistent with negligible dust attenuation.

Complicating this however is the fact that the UV continuum is also affected by other properties intrinsic to the stellar population, including the previous star formation history, metallicity, initial mass function and the presence of Lyman-$\alpha$ emission (unless it is outside the range of the UV continuum probed). While the sensitivity of the UV continuum colour to different influences complicates its interpretation as an estimate of dust attentuation, it may provide a technique to probe a range of other physical properties, especially at high redshift (e.g. Bouwens et al. 2010a). 

High-redshift galaxies are typically selected using a form of the Lyman-break technique (e.g. Steidel, Pettini and Hamilton 1995, Steidel et al. 1999) in which galaxies are principally selected on the basis of an extremely red colour, the cause of which is interpreted as the result of the Lyman-$\alpha$ break. This is typically combined with an additional colour selection and morphological criteria to guard against intrinsically red contaminants, such as low-mass dwarf stars and lower-redshift evolved stellar populations (e.g. Stanway et al. 2008). However this makes studying the distribution of UV continuum colours difficult as most selection criteria designed to maximise the number of high-redshift candidates introduces both explicit and implicit biases. In this work we define a selection criterion which minimises these biases and use it to identify a catalogue of high-redshift star forming galaxies. Using this catalogue we investigate the rest-frame UV continuum colour distribution as a function of both luminosity and redshift. This paper is organised as follows: in Section 2 we describe the observations, data reduction and photometry. This is followed in Section 3 by a description of our selection criteria and how it was derived, together with simulations used to probe its effectiveness. In Section 4 we outline the various properties that affect the intrinsic UV continuum slope, and investigate how changing these properties (including the initial mass function and the star formation history) quantitatively affects the UV continuum colour. In Section 4 we present our measurements of the UV continuum colour, together with the colour excess $E(B-V)$ (inferred assuming the Calzetti et al. 2000 extinction curve) and the slope of the UV continuum $\beta$ (i.e. $f_{\lambda}\propto\lambda^{\beta}$) of our sample split by redshift and also luminosity. Finally in Section 5 we present our conclusions. Throughout this work we employ the $AB$ magnitude system (Oke \& Gunn 1983) and a $\Lambda$CDM cosmology ($\Omega_{\Lambda}=0.7$, $\Omega_{\rm M}=0.3$, and $H_{0}=70\,{\rm km\,s^{-1}\,Mpc^{-1}}$)

\section{Observations, Catalogue Construction and Photometry}

In this work WFC3 observations of three fields (HUDF, P34 and P12, see Wilkins et al. 2011) in the vicinity of the CDFS/GOODS-South are combined with existing Advanced Camera for Surveys (ACS) ($b_{435w}$ (unavailable for the P34 and P12 fields), $v_{606w}$, $i_{775w}$ and $z_{850lp}$) imaging. The WFC3 data come from the {\em HST} Treasury programme GO-11563 (P.I.\ G.~Illingworth) and consist of observations in three filters: $Y_{105w}$, $J_{125w}$ \& $H_{160w}$. A more detailed account of the observations, data reduction, and photometry is presented in Wilkins et al. (2011) and Bunker et al. (2010). In each of the three fields the $2\sigma$ point source detection limit is $>29.50$ (AB) for the ACS images and $>29.15$ for the WFC3 images with the exception of the $H_{160w}$ imaging of the P12 field ($\simeq 28.2$).

\section{Candidate Selection}\label{sec:selection}

The combination of the {\em HST} ACS and WFC3 imaging available allows us to identify star forming galaxies, and measure their rest-frame UV continuum colours from $z=2 \to 8$ (for the HUDF field). However, the addition of the new deep WFC3 NIR $Y_{105w}$, $J_{125w}$, and $H_{160w}$ imaging is most relevant to the highest redshifts ($z>4.5$) and it is these objects on which we focus our attention.

\subsection{Lyman-Break Technique}

At sufficiently high redshift an increasing number of absorption lines due to intervening neutral hydrogen causes a decrease in the flux shortward of the rest-frame location of Lyman-$\alpha$. At the lower redshift end this is manifested by the Lyman-$\alpha$ forest (Songaila \& Cowie 2002); at very-high redshift the large number of absorption lines effectively effectively extinguishes the flux shortward of $1216{\rm \AA}$, producing the Gunn-Peterson trough feature (Gunn \& Peterson 1965).

In either case, the flux measured through broadband photometry shortward of rest-frame Lyman-$\alpha$ will be reduced relative to that above (longward of) the break. This can be exploited by using at least two filters are positioned above and below the observed frame location of the Lyman-$\alpha$ break.  

\begin{figure}
\centering
\includegraphics[width=20pc]{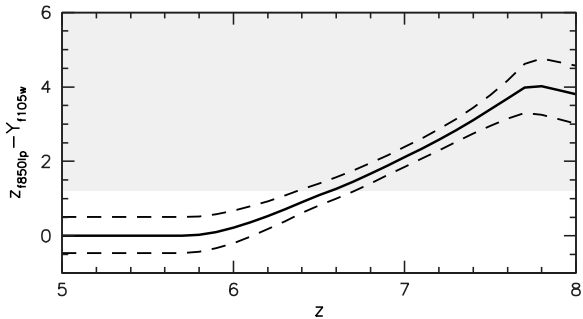}
\caption{The $(z_{f850lp}-Y_{f105w})_{AB}$ colour evolution of a galaxy assuming $f_{\lambda}=\lambda^{\beta}$ at $\lambda_{\rm rest} > 1216{\rm \AA}$, $f_{\lambda}=0.01\times \lambda^{-2.0}$ at $912{\rm \AA} <\lambda_{\rm rest} < 1216{\rm \AA}$ (i.e. intervening neutral Hydrogen absorbs $\sim 99\%$ of the emitted photons) and $f_{\lambda}=0.0$ at $\lambda_{\rm rest} < 912{\rm \AA}$. The three lines denote $\beta=1.0$, $-2.0$ and $-5.0$ from top to bottom and seek to highlight that the redshift range probed by single colour selection is sensitive to the underlying continuum slope.}
\label{fig:zc}
\end{figure}

Figure \ref{fig:zc} shows the $(z_{f850lp}-Y_{f105w})_{AB}$ colour of a galaxy with a simple, yet physically motivated spectrum ($f_{\lambda}=\lambda^{\beta}$ at $\lambda_{\rm rest} > 1216{\rm \AA}$ with $\beta=-2.0$ \footnote{An approximation to the slope expected for a young star forming population with no dust}, $f_{\lambda}=0.01\times \lambda^{-2.0}$ at $912{\rm \AA} <\lambda_{\rm rest} < 1216{\rm \AA}$ (i.e. intervening neutral Hydrogen absorbs $\sim 99\%$ of the emitted photons) and $f_{\lambda}=0.0$ at $\lambda_{\rm rest} < 912{\rm \AA}$) as a function of redshift. At $z>6.$ the $z_{f850lp}$ filter encompasses the Lyman-$\alpha$ break resulting in a decrease in the recorded flux and an increasingly red $(z_{f850lp}-Y_{f105w})_{AB}$ colour. Using a $(z_{f850lp}-Y_{f105w})_{AB}>1.0$ cut should then efficiently select galaxies at $z>6.5$. While galaxies at $z>8.0$ will have colours consistent with $(z_{f850lp}-Y_{f105w})_{AB}>1.0$ the $Y_{f105w}$-band filter is increasingly contaminated by the break reducing the recorded flux and thus the effective selection volume.

\subsection{Contamination}

While a single colour criteria is capable of discriminating high-redshift objects from the bulk of the galaxy population it is still liable to several types of contamination, including both objects with intrinsically red colours and transient phenomenon.

Intrinsically red objects predominantly include galaxies with prominent Balmer/$4000{\rm \AA}$ breaks located at lower redshifts than the target population (for example these galaxies at $z\simeq 1.5$ would contaminate a pure $z_{f850lp}-Y_{f105w}$ selected sample). Galaxies with prominent Balmer/$4000{\rm \AA}$ breaks however typically have intrinsically very red spectra at wavelengths longward of the break location, and can be idenitfied using additional colour information. Cool, low-mass dwarf stars are an additional contaminant (late $M$-class for the $v$-drop selected galaxies and $LT$-class for the $i$ and $z$-drops). While in some cases these can also be excluded on the basis of additional colour information, for certain filter combinations this is not possible and stars can only be excluded on morphological grounds. However this is problematic especially at low signal-to-noise and may also result in the removal of real high-redshift galaxies (see Stanway et al. 2008, Wilkins et al. 2011).

Transient phenomenon capable of contamination include stars with high-proper motions or supernovae. Transient phenomenon are potential contaminants when there is a significant gap between the acquisition of images in neighbouring bands. In the context of the observations used in this work this occurs between the {\rm ACS} $bviz$-band and {\rm WFC3} $YJH$-band imaging. An object with an observed {\em very}-red $z_{f850lp}-Y_{f105w}$ colour (i.e. no detection in the $z_{f850lp}$-band) may potentially be a transient phenomenon. While the brightest transients can be excluded by examining imaging at a previous epoch (assuming it has roughly the same spectral coverage) fainter objects are less easy to unambiguously identify.

\subsection{Probing the Rest-Frame UV Continuum of high-$z$ Galaxies}

To measure the rest-frame UV continuum colours without the possibility of contamination from either the Lyman-$\alpha$ break or line emission at least two filters above the break are necessary - subsequently we refer to these two filters as $C$ and $D$ such that $\lambda_{D}>\lambda_{C}$. In the context of the available WFC3 imaging, the requirement of two filters safely above the Lyman-$\alpha$ break limits us to galaxies selected as $v_{f606w}$, $i_{f775w}$ and $z_{f850lp}$ dropouts (objects with red $(A-B)$ colours, where the $A$ filter is the drop-out filter and the $B$ filter is the next available filter moving redward); covering approximately $4.7<z<8.0$. While it is also possible to identify even higher redshift $Y_{f105w}$ and $J_{f125w}$ dropout galaxies (e.g. Bouwens et al. 2010, Bunker et al. 2010, McClure et al. 2010, Lorenzoni et al. 2011) there is insufficiently deep imaging in bands above the rest-frame Lyman-$\alpha$ break to measure the UV continuum properties without the possibility of contamination from Lyman-$\alpha$ emission or the break itself. 

\begin{figure}
\centering
\includegraphics[width=20pc]{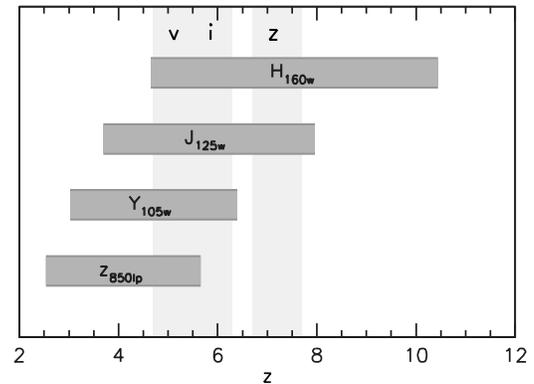}
\caption{Horizontal bars denote the redshift range over which each filter probes the rest-frame UV continuum from $1250-3000\,{\rm \AA}$. The vertical bars denote the rough redshift range for each dropout sample.}
\label{fig:z_range}
\end{figure}

In Figure \ref{fig:z_range} we show the redshift-range over each of the three WFC3 filters and the ACS $z_{f850lp}$ can be used to probe the rest-frame UV continuum from $1250$\footnote{$1250{\rm\AA}$ is chosen to provide sufficient distance from Lyman-$\alpha$.} to $3000{\rm \AA}$. Where the ranges of two filters overlap they can be used to {\em safely} probe the UV continuum free from contamination due to the Lyman-$\alpha$ break or Lyman-$\alpha$ line emission. At $z\simeq 7$ (where $z$-drop selected galaxies will lie) only the $J_{f125w}$ and $H_{f160w}$ filters lie safely above the Lyman-$\alpha$ break and it is these filters that we adopt to probe the UV continuum slope. At $z\simeq 6$ the $Y_{f105w}$, $J_{f125w}$ and $H_{f160w}$ WFC3 filters all lie within the restframe $1250-3000{\rm \AA}$ range and can thus be used to probe the UV continuum - however to aid in the comparison between different redshifts we utilise only the $Y_{f105w}$ and $J_{f125w}$ filters which cover a similar rest-frame wavelength range to the $J_{f125w}$ and $H_{f160w}$ filters at $z\simeq 7$. At $z>8$ the $J_{f125w}$ filter extends shortward of the rest-frame Lyman-$\alpha$ and thus cannot be robustly used to measure the continuum colour due to the possibility of contamination from the break and/or Lyman-$\alpha$ line emission. 

The redshift range over which individual filters can be used to safely probe the rest-frame UV continuum has important implications for the specific selection criteria used; this is discussed in more detail in the following section. 

\subsection{Selection Criteria}

\begin{figure}
\centering
\includegraphics[width=16pc]{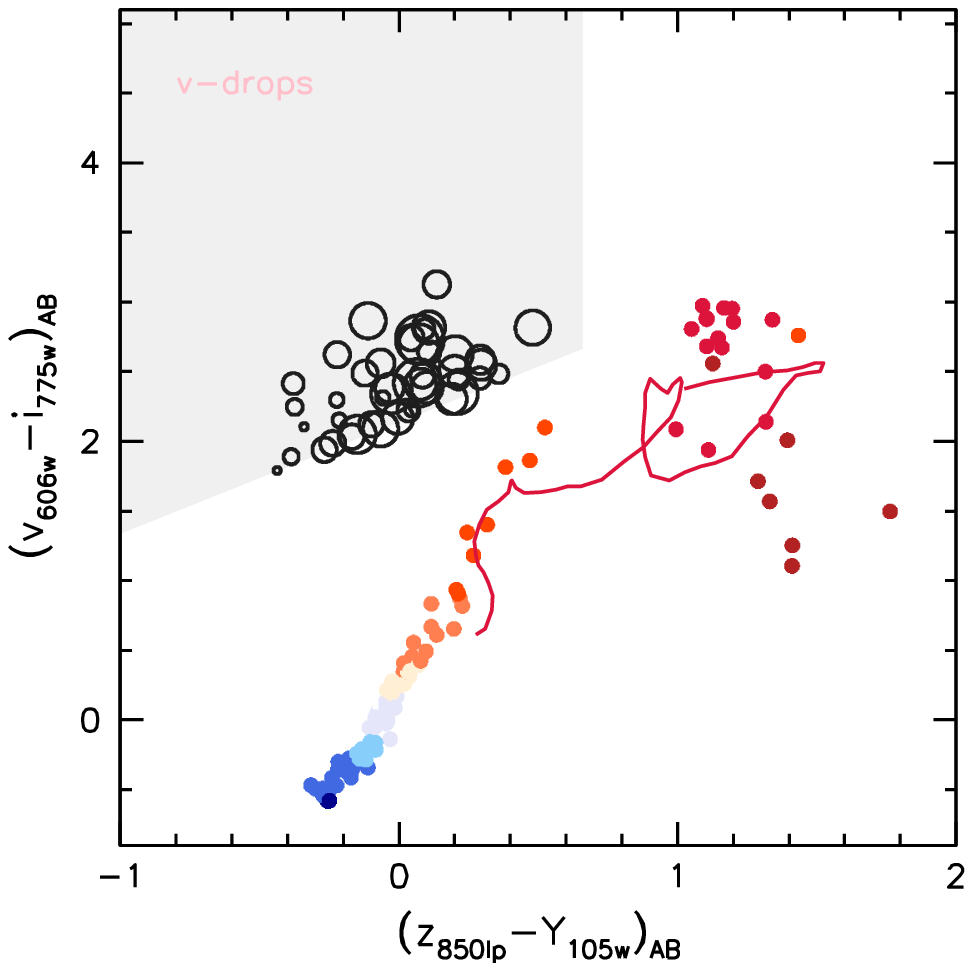}
\includegraphics[width=16pc]{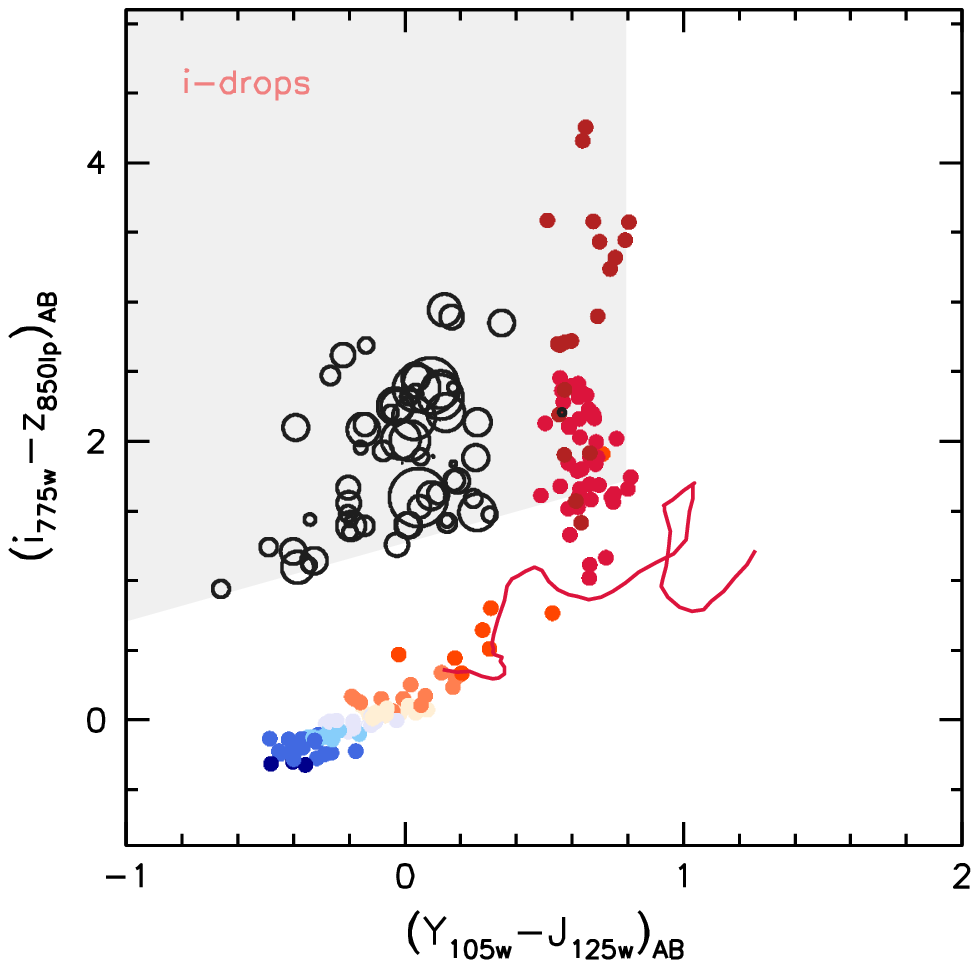}
\includegraphics[width=16pc]{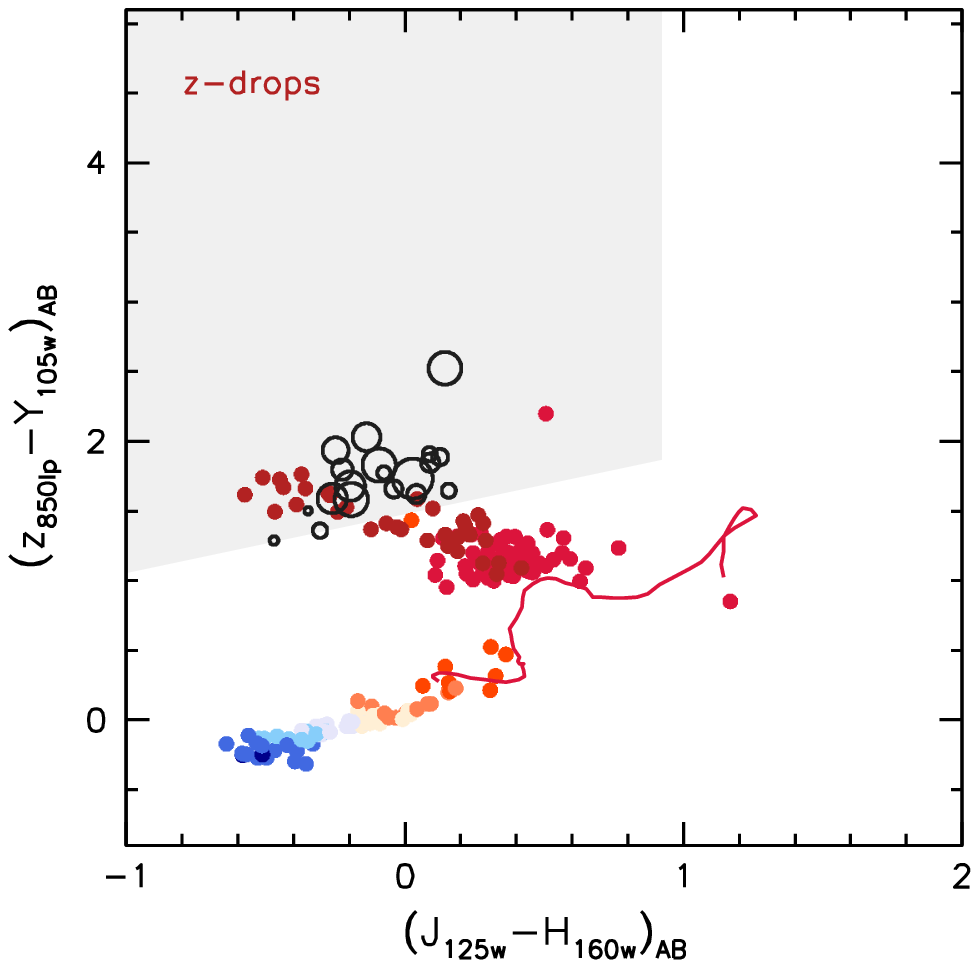}
\caption{$(C-D) - (A-B)$ colour-colour (where the $(A-B)$ colour probes the Lyman-$\alpha$ break, e.g. for the $z_{f850lp}$-drops $(A-B)\equiv (z_{f850lp}-Y_{f105w})$ and $(C-D)$ colour is used to probe the UV continuum, see Table \ref{tab:params}) figures for each of the dropout samples ($v$, $i$ and $z$ from the top respectively) showing the selection window (grey), the positions of $O\to T$ class dwarf stars, the track taken (over a narrowed redshift range where the break colour probes the Balmer/$4000{\rm \AA}$ break) by an instantaneous burst of star formation at $z=10$, and the location of our candidate objects. The size of the symbol is an indication of the absolute magnitude $M_{1500}$ and error bars are omitted for clarity.}
\label{fig:cc}
\end{figure}

\begin{table*}
\begin{tabular}{ccccccccccccc}
drop & break-colour$^{a}$ & slope-colour &$\langle z \rangle$$^{b}$ & $z$-range & rest-frame &$N_{obj}$$^{d}$ & \multicolumn{5}{|c|}{selection parameters}\\
$A$ & $(A-B)$ & $(C-D)$ &  &  & $C_{\lambda}$ at $\langle z \rangle$$^{c}$  &  & $p_{1}$ & $p_{2}$ &$p_{3}$ & $p_{4}$ & $p_{5}$ \\
\hline
$v_{f606w}$ & $(v_{f606w}-i_{f775w})$ & $(z_{f850lp}-Y_{f105w})$ & $5.0$ & $4.7-5.3$ &  $1416{\rm \AA}$ & $34$ & $0.62$ &$1.48$ & $0.42$ & $0.89$ & $0.46$\\
$i_{f775w}$ & $(i_{f775w}-z_{f850lp})$ & $(Y_{f105w}-J_{f125w})$ & $5.9$ & $5.5-6.3$ &  $1522{\rm \AA}$ & $52$ & $0.78$ &$1.27$ & $0.56$ & $0.73$ & $0.66$\\
$z_{f850lp}$ & $(z_{f850lp}-Y_{f105w})$ & $(J_{f125w}-H_{f160w})$ & $7.2$ & $6.7-7.7$ &  $1524{\rm \AA}$ & $18$ & $0.95$ &$2.14$ & $0.08$ & $0.83$ & $0.76$\\
\end{tabular}
\caption{Summary of the choice of $ABCD$ filters, where the $(A-B)$ colour probes the Lyman-$\alpha$ break, e.g. for the $z_{f850lp}$-drops $(A-B)\equiv (z_{f850lp}-Y_{f105w})$ and $(C-D)$ colour is used to probe the UV continuum, redshift range, and the selection parameters employed for each drop out sample. $^{a}$Filter combination used to probe the Lyman-$\alpha$ break. $^{b}$Assumes no evolution within the redshift range. $^{c}$The rest-frame central wavelength of the $C$ filter assuming the source is at $z=\langle z \rangle$. $^{d}$Number of objects meeting the selection criteria in all three fields. }
\label{tab:params}
\end{table*}

While a simple selection criteria based solely on the break colour $(A-B)$ can be used to identify high-redshift galaxies it is both susceptible to contamination and can introduce a number of biases when attempting to study the rest-frame UV properties of high-redshift galaxies. Each of these factors much be taken into account when designing a selection criteria. 

Firstly, contamination from {\em lower}-redshift objects has to be minimised. Figure \ref{fig:cc} shows the positions of $O\to T$ class dwarf stars (filled points) and the track followed by a stellar population formed from through an instantaneous burst at $z=10.0$ (maximising the strength of the $4000{\rm \AA}$ feature at $z=1.5$). In order to exclude these objects, while still maintaining a reasonable number of objects we introduce a cut on the colour used to probe the UV continuum ($(C-D)$, see Table \ref{tab:params}), i.e. $(C-D)<p_{1}$. The precise value of $p_{1}$ is chosen such that for each dropout sample we probe to the same maximum value of $E(B-V)_{Calzetti}=0.8$. The $(C-D)$ colour corresponding to this value of $E(B-V)$ is determined by applying the Calzetti et al. (2000) reddening curve to a synthetic SED assuming $100\,{\rm Myr}$ previous constant star formation, solar metallicity and a Salpeter (1955) IMF. This is further elaborated on in Section 4 where we discuss the impact of various effects on the UV continuum slope of star forming galaxies. The value of $p_{1}$ for each drop-sample (together with the other selection criteria parameters) is presented in in Table \ref{tab:params}. It is important to note that such a criterion clearly imposes an explicit bias against very dusty galaxies, however as we will see in Section \ref{sec:results} this appears to be only problematic if the colour/dust distributions are bimodal. To further guard against potential contamination we also introduce a criterion that candidate high-$z$ objects not have a detection in any filter which probes rest-frame wavelengths of $<912{\rm\AA}$ (below the Lyman limit).  

The second consideration is to ensure that the redshift range (and mean redshift) probed is not sensitive to the absolute luminosity or intrinsic UV continuum colour, and further, that for a given absolute luminosity the effective volume survey is as {\em uniform} as possible. A related consideration is that the theoretical redshift range should be compatible with the range over which the available filters can be used to safely probe the UV continuum. For example, as noted in the previous section, it is necessary to restrict the $z$-drop redshift range to $z<8.0$ so that the $J_{f125w}$ is not contaminated by either Lyman-$\alpha$ emission or the break. 

The first step in this procedure is to determine the relationship between the break colour ($A-B$) at a given redshift and the uncontaminated UV continuum slope colour ($C-D$). This is necessary as objects with different rest-frame UV colours will, as a function of their absolute magnitude defined at some arbitrary point on the rest-frame UV continuum, enter a single colour selection at a different redshift. To understand this more clearly, it is useful to consider the case of two galaxies with the same rest-frame absolute magnitude measured at $1500{\rm \AA}$ ($M_{1500}$) but different spectral slopes; one with a red slope ($\beta=0.0$, where $f_{\lambda}=\lambda^{\beta}$) and the second with a blue slope ($\beta=-2.0$). If both galaxies have a sharp break at $\lambda=1216{\rm \AA}$, blueward of which the flux density is reduced to $1\%$ of that assuming no break, filters located to straddle the break would record different colours for each galaxy. If candidates were selected solely on the basis of the single colour, the redshift range for galaxies with one spectral slope would differ from that for another, potentially resulting in a continuum colour bias. This can be seen in Figure \ref{fig:zc} where the $z_{f850lp}-Y_{f105w}$ colour evolution track is shown assuming three values of $\beta$ ($\beta \in \{1.0,-2.0,-5.0\}$). Each track crosses a given $z_{f850lp}-Y_{f105w}$ value at a different redshift.

To account for this effect we introduce a criterion that relates the break colour $(A-B)$ to the observed colour used to probe the continuum $(C-D)$ at a given redshift, i.e. a criterion which requires that the  $(A-B)>f(C-D)_{z}$. To ascertain this relationship we use a range of UV continuum slopes and determine $(A-B)$ as a function of the $(C-D)$ for a given redshift. This redshift ultimately forms the lower-boundary of the selection range and is chosen such that the selection window is sufficiently removed from the majority of the contaminant population. The relationship between $(A-B)$ and $(C-D)$ is found to be approximately linear thus we introduce a criteria: $(A-B)>p_{2}\times (C-D)+p_{3}$. The selection window in $(A-B)-(C-D)$ colour space resulting from both this criteria and with the $(C-D)<p_{1}$ criteria is shown in Figure \ref{fig:cc}. The shape of the selection window has the additional benefit in that for a given redshift boundary it reduces the amount of contamination compared to assuming a simple $(A-B)$ based criteria. Indeed, the colour criteria often employed to optimise the number of robust candidates typically has a similar shape. 

While these two criteria can be used to effectively fix the lower limit of the redshift range, the upper limit remains sensitive to the UV continuum colour, and for the $z_{f850lp}$ drop galaxies is above the redshift at which the $J_{f125w}$ filter can be used safely. The solution to this problem is to introduce a further colour criteria, this time using the $(B-C)$ and $(C-D)$ colours (i.e. $(B-C) < p_{4}\times (C-D)+p_{5}$). The parameters $p_{4}$ and $p_{5}$ are determined in a similar technique to $p_{2}$ and $p_{3}$.

In addition to the colour criterion outlined above we also introduce an abolsute magnitude $M_{1500}$ criteria such that $M_{1500}<-18.5$. $M_{1500}$ is determined using the absolute magnitude inferred from the apparent $C$ magnitude assuming the source lies at the mean redshift and applying a correction using the $(C-D)$ colour to scale it to $1500{\rm\AA}$. Given that the pivot wavelength of the $C$-filter at the mean redshift of each drop-out sample is $\simeq 1500{\rm\AA}$ this correction is very small. For this criteria we choose $M_{1500}<-18.5$ corresponding to $z_{f850lp}=27.89$, $Y_{f105w}=28.1$, $J_{f125w}=28.5$ at $z=5.0$, $z=5.9$ and $z=7.2$ respectively. For the $v_{f606w}$ and $i_{f775w}$ selected drops this is significantly brighter than the $5\sigma$ point source limit, however for the $z_{f850lp}$-drop selection this limit is close to the $5\sigma$ point source limit suggesting large uncertainties and incompleteness. Indeed, as will be shown in the proceeding section, where we employ a series of photometric scatter simulation this appears to be the case, and reference is made to this when we analyse our results.

In summary, the selection criteria we employ can be defined as:
\begin{eqnarray*} 
M_{1500}&<&-18.5 \\
(C-D) & < & p_{1}  \quad \equiv \quad (E(B-V)_{\rm Calzetti}<0.8)   \\
(A-B) & > & p_{2}\times (C-D)+p_{3} \\
(B-C) & < & p_{4}\times (C-D)+p_{5} \\
\end{eqnarray*} 
together with a non-detection ($<2\sigma$) condition in the relevant optical filters. These filters, and the selection criteria parameters ($p_{i}$) are given for each dropout sample in Table \ref{tab:params}. 

To verify our selection criteria and investigate any further biases we conduct a series of photometric simulations; these are described in the following section. 

Finally, it is important to stress that these selection criteria are designed to minimise any potential biases in the measurement of the UV continuum slope. A result of this is that the number of candidates selected using this criteria is potentially much smaller than a criteria designed solely to maximise the number of high-redshift galaxy candidates.

\subsection{Simulated Redshift Distributions and Effective Volumes}\label{sec:effvol}

The presence of photometric noise will have the effect of scattering objects both in and out of the selection window, possibly distorting the redshift distribution and affecting the effective volume. Such effects may introduce an additional colour bias (other than that explicitly defined in the selection criteria, i.e. $E(B-V)_{\rm Calzetti}<0.8$) or affect the mean redshift.

\begin{figure}
\centering
\includegraphics[width=20pc]{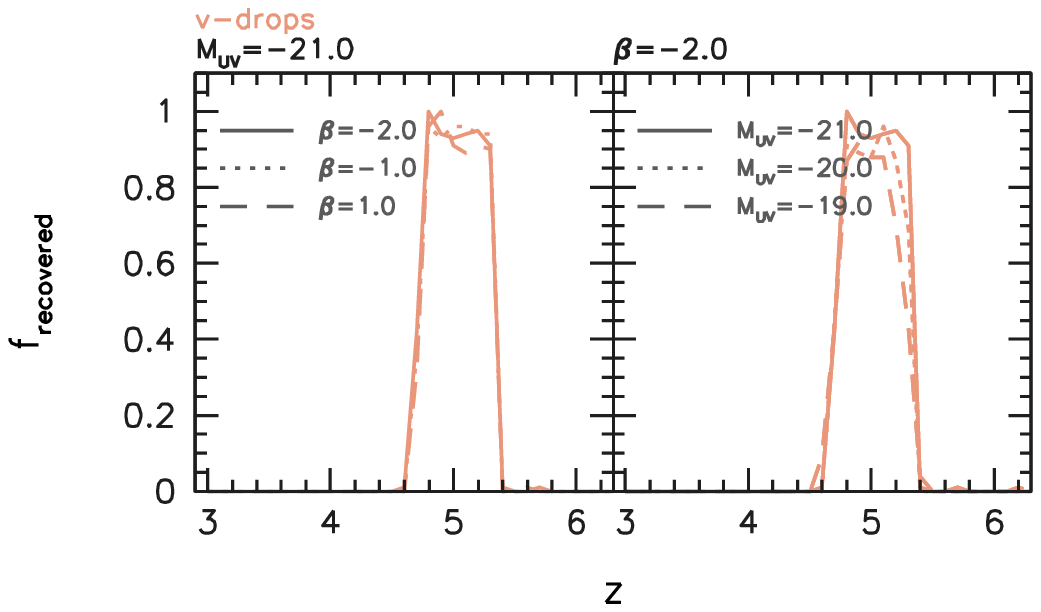}
\includegraphics[width=20pc]{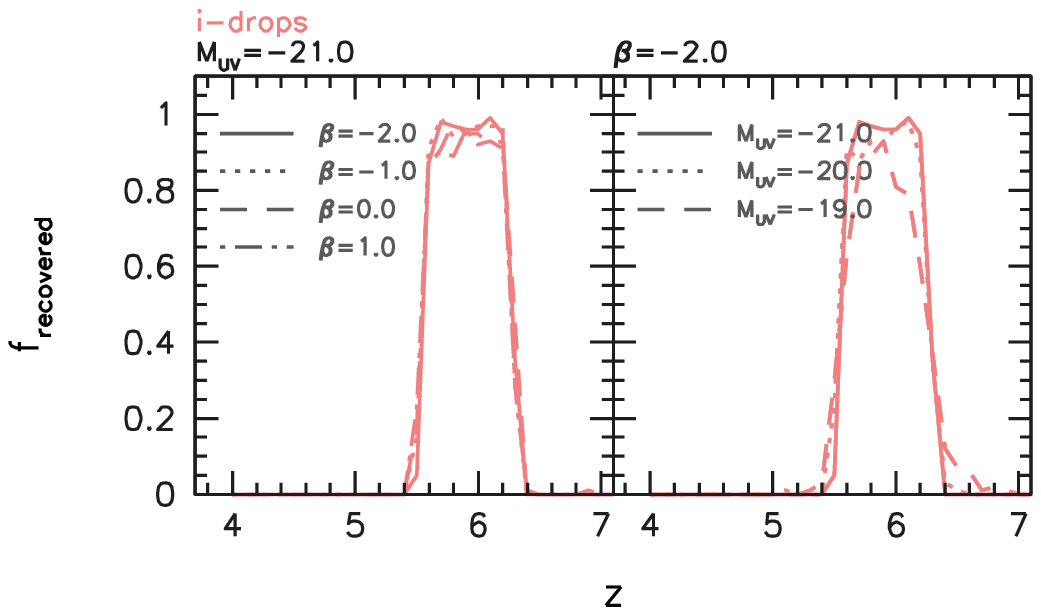}
\includegraphics[width=20pc]{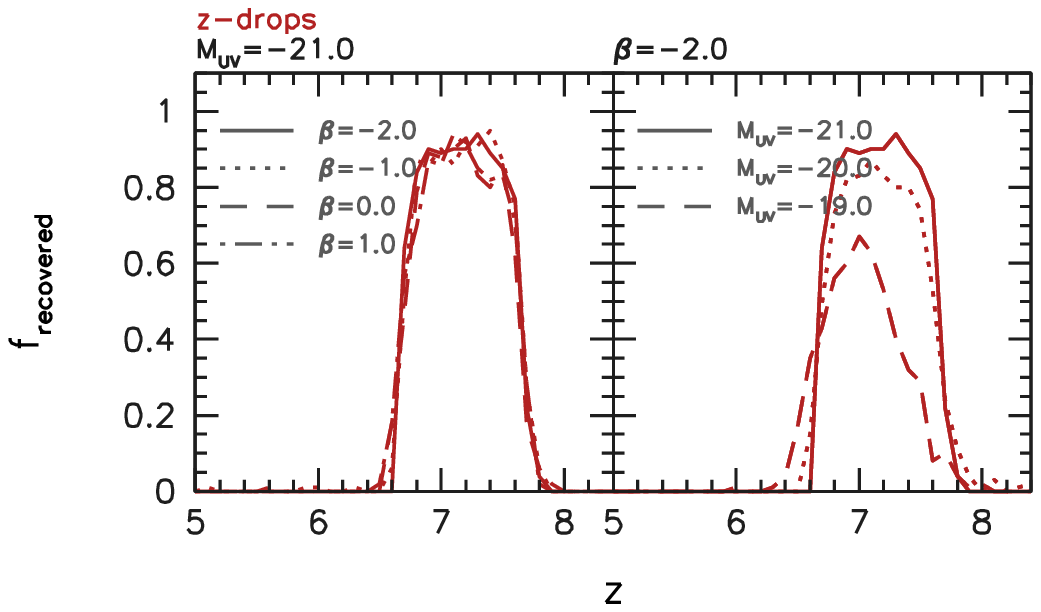}
\caption{The simulated probability of recovering a synthentic galaxy inserted into the original images (i.e. {\em recoverability}) as a function of redshift assuming different properties. {\em left panel} Assuming $M_{1500}=-21.0$ and various values ($\beta \in \{ -2.0,-1.0,0.0,1.0\} $) of the intrinsic UV continuum slope $\beta$. {\em right panel}  Assuming $\beta=-2.0$ and various values ($M_{1500} \in \{ -21,-20,-19\} $) of the rest-frame UV luminosity.}
\label{fig:rec_z}
\end{figure}

To investigate this we perform a series of simulations in which we populate the original images with synthetic sources with various properties (common properties including: IGM absorption as given by Madau et al. 1996, and sizes and profiles consistent with the secure candidates from Wilkins et al. 2011). Using the same techniques (i.e. use of {\sc SExtractor}, aperture photometry, and selection criteria) that are used to identify candidate high-$z$ objects we assess the ability to recover objects based on their UV colours, redshifts and absolute magnitudes. An example of this is shown in Figure \ref{fig:rec_z}; this figure shows the {\em recoverability}, i.e. the fraction of sources inserted that are recovered using our selection criteria, as a function of redshift for various rest-frame UV absolute magnitudes ($M_{1500}\in\{-21.0,-20.0,-19.0\}$) and UV continuums slopes ($\beta\in\{-2.0,-1.0,0.0,1.0\}$). 

For the majority of cases the shape of the recovered redshift distribution (and importantly the mean redshift) closely matches that which we defined previously, further, while the average recovered fraction decreases at fainter luminosities and redder UV continuum slopes this effect is mild with a single exception. In the case of the faintest $z$-drops case we see that both the recovered redshift distribution is changed (with the mean recovered redshift now lower than $z=7.2$) and the average recovered fraction is decreased by $\simeq 50$ percent.

The effect of the rest-frame UV absolute magnitude and colour in affecting redshift distribution and mean recoverability can be shown more clearly by estimating the effective volume for each combination of parameters. To allow easier comparison we normalise this to the volume if the entire theoretical redshift range (as defined by the selection criteria) is probed uniformly with a probability of recovery of $100$ percent. 

\begin{figure}
\centering
\includegraphics[width=20pc]{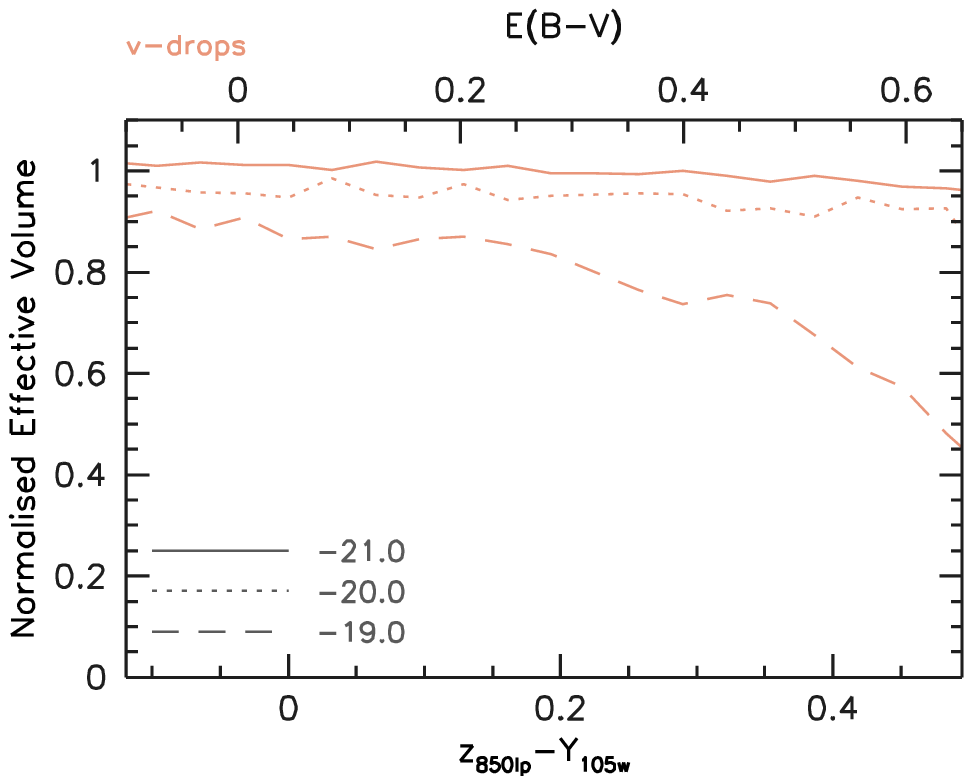}
\includegraphics[width=20pc]{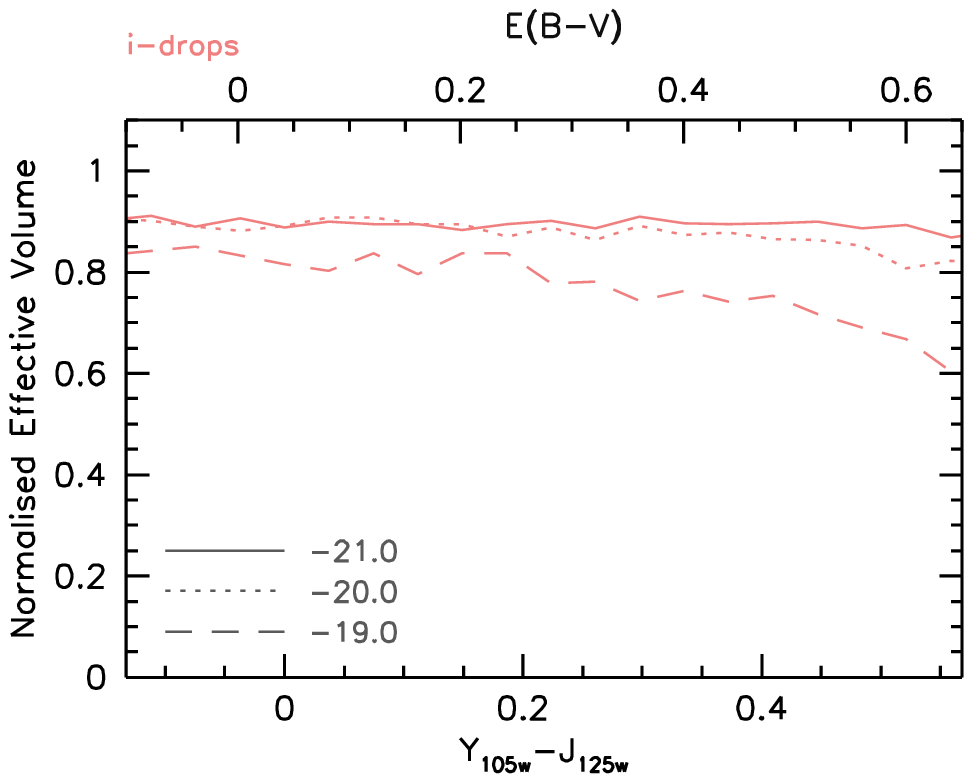}
\includegraphics[width=20pc]{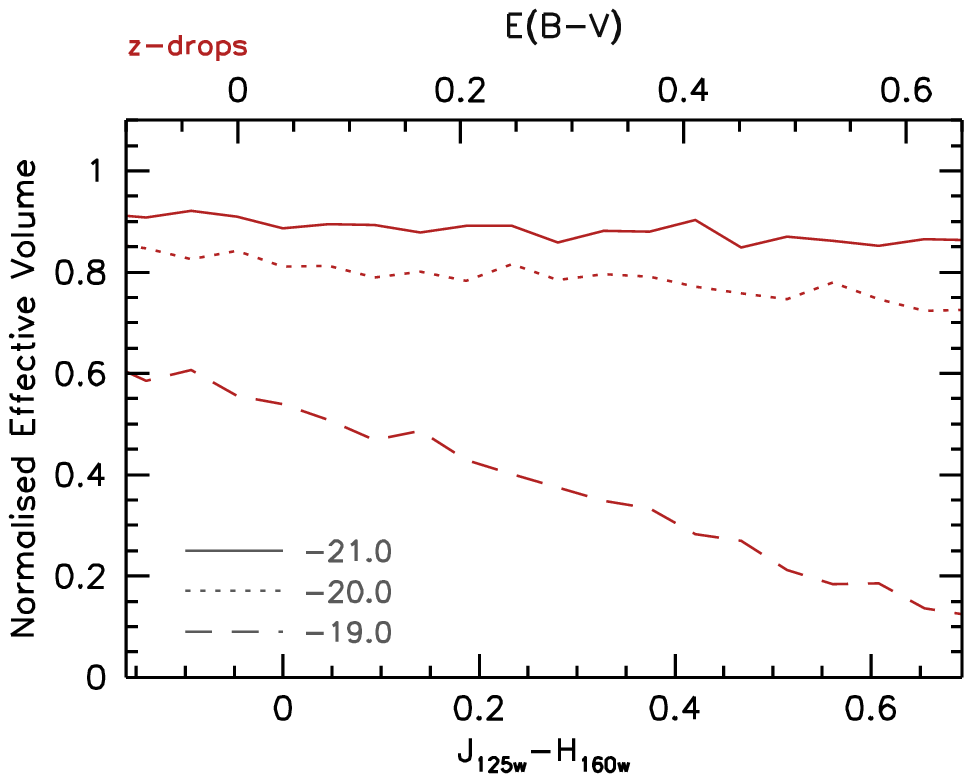}
\caption{The {\em relative} (to the maximum achievable) effective volume probed as a function of the UV continuum slope for different absolute magnitudes for each drop sample for the combination of HUDF, P12 and P34.}
\label{fig:vol}
\end{figure}

The normalised effective volume is shown, as a function of intrinsic UV colour, for the three dropout criteria and assuming three different absolute magnitudes in Figure \ref{fig:vol}. In the brightest two bins there is only a $< 10\%$ decrease in the effective volume over the range of colours we probe. For the faintest magnitudes considered, and in particular the highest-redshift $z$-drops, the effective volume falls significantly (for the $z$-drops the normalised effective volume is: $\approx 0.55$ assuming $(J_{f125w}-H_{f160w})_{AB}=0.0$). For these faintest magnitudes there is also a significant bias against objects with red rest-frame UV colours; for example the normalised effective volume of $z$-drops with $M_{1500}=-19.0$ and $(J_{f125w}-H_{f160w})_{AB}=0.6$ is $\approx 0.20$ (c.f.  $(J_{f125w}-H_{f160w})_{AB}=0.0$: $\approx 0.55$). When examining the observed distribution of UV colours it is important to take these biases into account. 

In Section \ref{sec:res.int}, where we consider the observed distribution of UV continuum colours we also use make use photometric scatter simulations in the context of ascertaining the scatter in the UV continuum colours we would expect purely due to photometric scatter.

\section{The UV Spectral Slope as a Diagnostic of Star Forming Galaxies}\label{sec:uv}

The UV continuum of star forming galaxies is dominated by emission from the most massive, high-temperature ($>3\,{\rm M_{\odot}}$, $OB$ class) stars. The UV ($1216\to 3000{\rm \AA}$) spectral flux density of the hottest of these stars is well characterised by a power law with a {\em blue} slope (i.e. if $f_{\lambda}\propto \lambda^{\beta}$ then $\beta<0.0$). This power-law behaviour principally arises because the UV continuum (from $1216\to 3000{\rm \AA}$) probes the Rayleigh-Jeans tail of the black body spectral energy distribution (though it deviates from a pure Rayleigh-Jeans behaviour because of the effects of opacity in the star's atmosphere). At lower effective temperatures the range $1216\to 3000{\rm \AA}$ probes the black body outside the Rayleigh-Jeans tail toward where the black body distribution peaks. This effect, combined with variation in the opacity (as a function of temperature etc.), results in the SED of the cooler, lower-mass stars having a redder UV continuum than that of stars with higher temperatures/masses. 

The sensitivity of the UV continuum of a star to its surface temperature (and opacity), and thus metallicity and mass means that the continuum of a composite stellar population is then sensitive to both the distribution of the masses of the stars and the metallicity. The distribution of masses is in turn determined by star formation history (SFH) and initial mass function (IMF); thus both these factors affect the observed UV continuum colours. In addition {\em extrinsic} properties such as the amount and composition of any gas and dust (either in the host galaxy or in intervening systems) can produce an effect. In particular the strong wavelength sensitivity of typical dust reddening curves can cause potentially large deviations in the UV continuum colours of galaxies.

In this section we investigate the magnitude of each of these effects. To aid with the comparison of different effects we define a {\em default scenario} about which we consider deviations in the amount of dust, the previous SFH, IMF and metallicity. This default scenario assumes $100\,{\rm Myr}$ continuous star formation, solar metallicity ($Z=0.02$), an IMF described by a two-part broken power with a Salpeter (1955) high-mass ($0.5<M/{\rm M_{\odot}}<120$) slope ($\alpha_{2}=-2.35$) and a shallower low-mass ($0.1<M/{\rm M_{\odot}}<0.5$) slope ($\alpha_{1}=-1.30$), no dust and is constructed using the {\sc Pegase.2} (Fioc and Rocca-Volmerange 1997, Fioc and Rocca-Volmerange 1999) population synthesis model. For a galaxy at $z=7.2$ (the mean redshift of the $z_{f850lp}$-drop sample) this scenario suggests a $(J_{f125w}-H_{f160w})$ colour of $-0.037$.

\subsubsection{The $\beta$ parameter}

To compare observations at different redshifts and/or using different filter combinations many studies parameterise the UV continuum in terms of index of the power law slope index $\beta$ (where $f_{\lambda}\propto \lambda^{\beta}$), where the default scenario suggests $\beta_{\rm default}\simeq -2.16$. However, some studies differ in their conversion from the observed colour to $\beta$; in this work we convert the $(C-D)$ colour to $\beta$ by determining the value of $\beta$ for a given colour assuming the underlying slope is perfectly represented by a power law. By this definition $\beta=s_{1}\times (C-D)-2.0$ where $s_{1}$ is equal to $6.21$, $5.37$ and $4.28$ for the $v$, $i$ and $z$-drops respectively. 

\subsection{Effects Intrinsic to the Stellar Population}

\subsubsection{Initial Mass Function}

As noted, the UV continuum of a star is sensitive to its initial mass, with more massive (hotter) stars (e.g. $O$ class) typically having bluer spectral slopes than the cooler ($B$ and $A$ class). Clearly then, any factor affecting the distribution of masses potentially can affect the slope of the composite stellar population. 

For example, a stellar population containing only stars with masses $>20\,{\rm M_{\odot}}$ (and thus containing only $O$ stars) will have a bluer UV continuum than a population including stars down to $0.1\,{\rm M_{\odot}}$. This can be seen in the top panel of Figure \ref{fig:C} where the $J_{f125w}-H_{f160w}$ colour of a stellar population at $z=7.2$ is shown as a function of the low mass cutoff ($m_{\rm low}$). For a population containing only the most massive $>100\,{\rm M_{\odot}}$ stars the predicted $J_{f125w}-H_{f160w}$ colour is $\approx -0.15$ ($\beta=-2.65$). 

The relative number of stars can also be affected by assuming an IMF with a different high-mass slope $\alpha_{2}$ (c.f. for a Salpeter IMF $\alpha_{2}=-2.35$). Increasing $\alpha_{2}$, i.e. flattening the slope, increases the contribution from the most massive stars again resulting in a blueward shift in the UV continuum colour. For example, increasing $\alpha_{2}$ to $-1$ (i.e. a uniform stellar mass distribution) results in $J_{f125w}-H_{f160w}\approx -0.09$ ($\beta=-2.39$). The sensitivity of the $J_{f125w}-H_{f160w}$ colour to $\alpha_{2}$ can be seen in the second panel of Figure \ref{fig:C}. 

\subsubsection{Star Formation History}

The strong initial mass dependence of the main-sequence lifetimes of stars means that the present day mass distribution, and thus UV continuum slope, is sensitive to the star formation history. After $<10\,{\rm Myr}$ of continuous star formation a population would would still retain {\em most} of its original high-mass stars. However, after prolonged periods of star formation some fraction of the most-massive stars will have evolved off the main-sequence. This reduces the relative contribution of these stars to the UV continuum resulting in a redder slope. In the third panel of Figure \ref{fig:C} the $(J_{f125w}-H_{f160w})$ colour (of a population at $z=7.2$) is shown as a function of the previous duration of continuous star formation. As the duration of previous star formation increases the UV colour becomes redder. After $10\,{\rm Myr}$ $(J_{f125w}-H_{f160w})\approx -0.07$ while for longer durations, $>1\,{\rm Gyr}$, $(J_{f125w}-H_{f160w})>0.0$. 

Star formation histories which are more stochastic, comprised of bursts and troughs, are capable of producing much redder UV continuum colours. For example, an instantaneous burst of star formation followed by $\sim 50\,{\rm Myr}$ of quiescence produces $(J_{f125w}-H_{f160w})\approx 0.07$, while still producing a significant UV luminosity. 

\subsubsection{Metallicity}\label{sec:interp.dust}

\begin{figure}
\centering
\includegraphics[width=19pc]{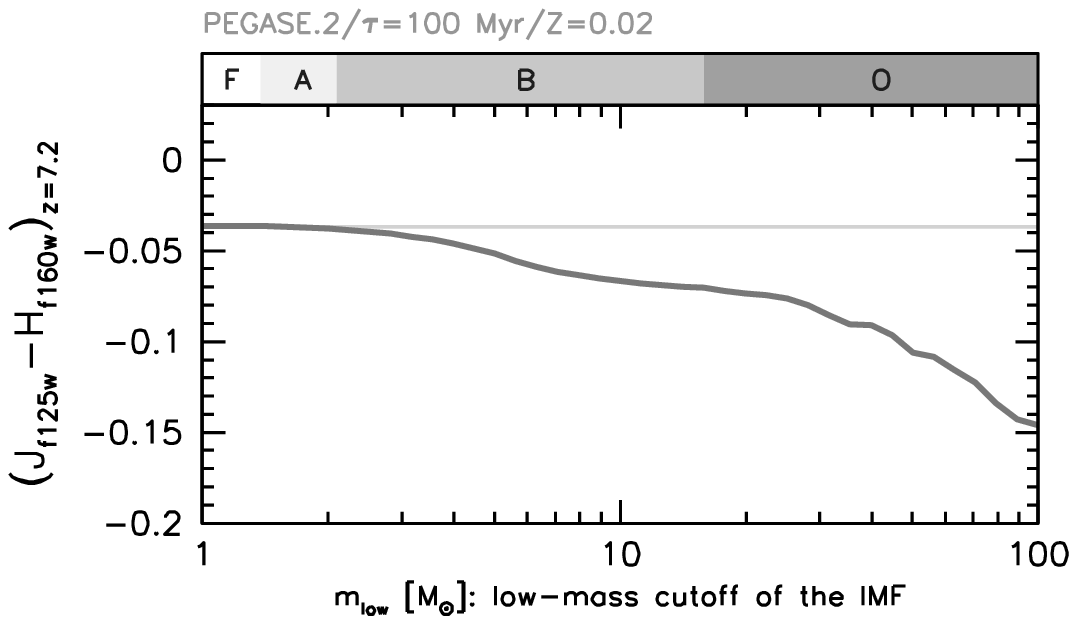}
\includegraphics[width=19pc]{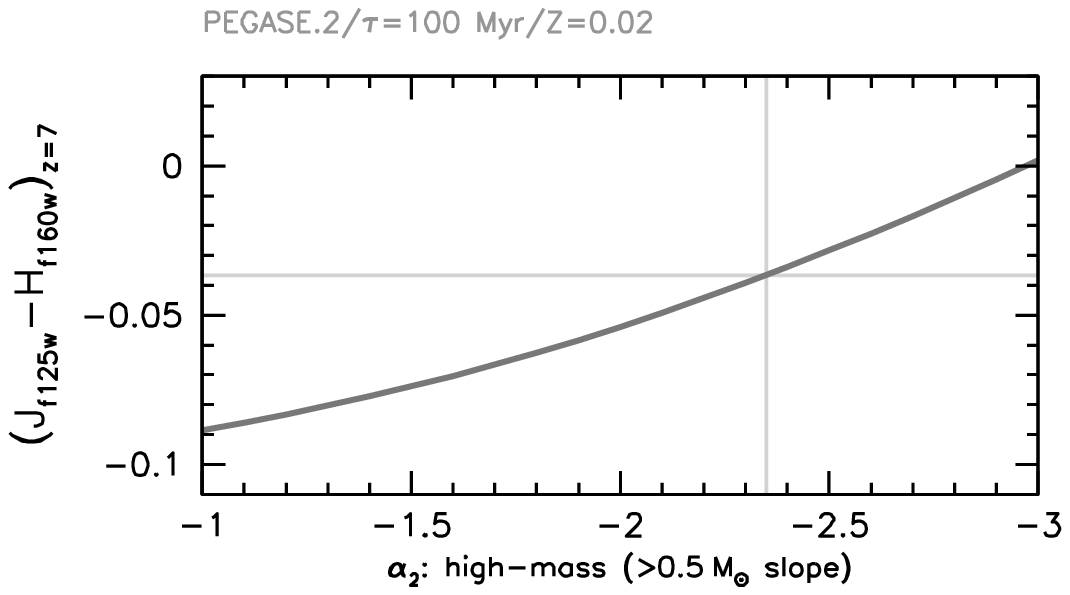}
\includegraphics[width=19pc]{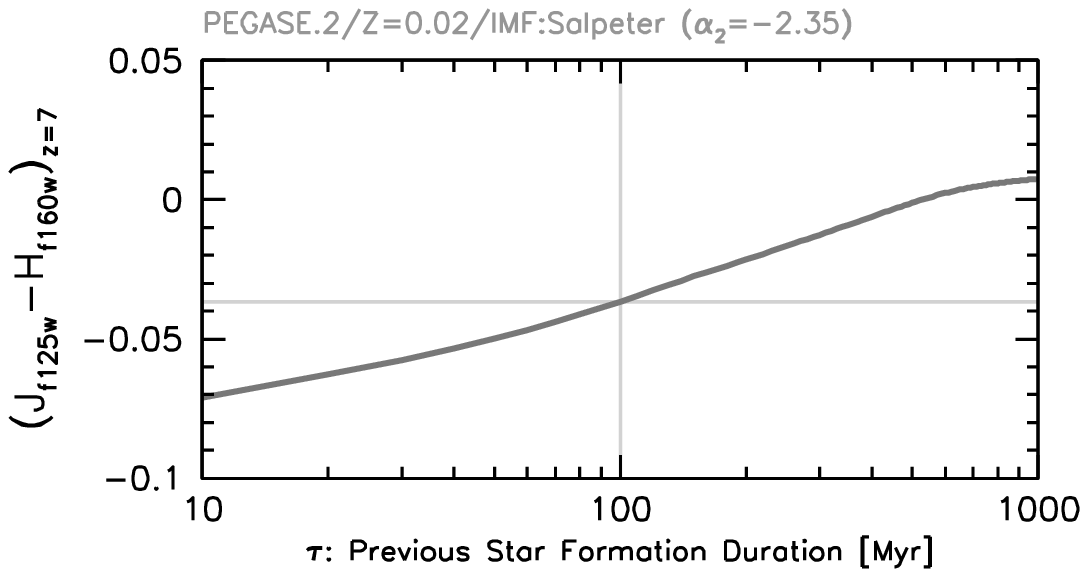}
\includegraphics[width=19pc]{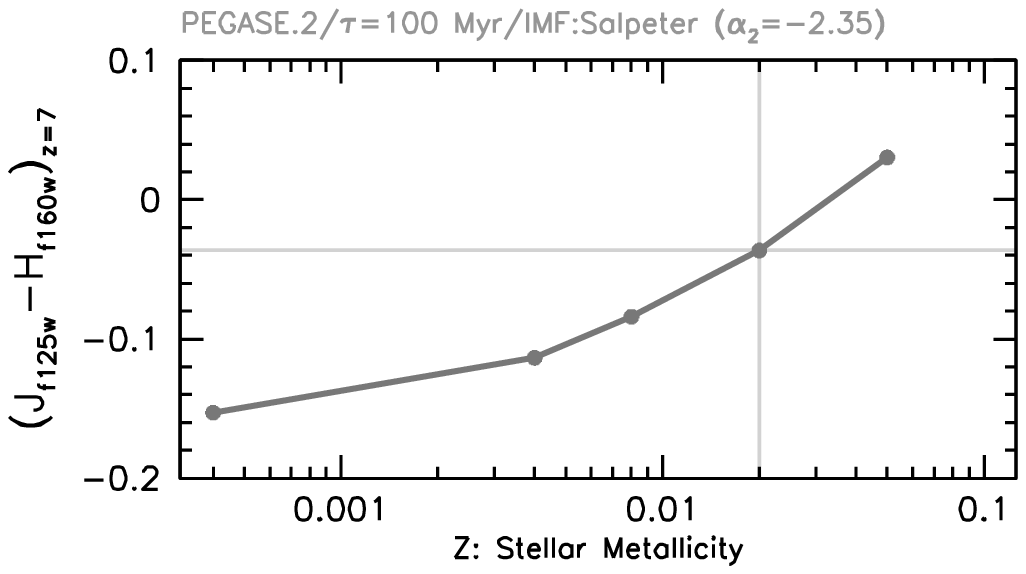}
\caption{The effect of changes to the lower stellar mass cutoff of the IMF ({\em top} panel), high-mass slope of the IMF ({\em 2nd} panel), duration of previous continuous star formation ({\em 3rd} panel) and metallicity ({\em bottom} panel) to the $J_{f125w}-H_{f160w}$ colour of a galaxy at $z=7.2$ from the {\sc Pegase.2} population synthesis model. The vertical and horizonal lines denote the value and colour for the {\em default} scenario respectively.}
\label{fig:C}
\end{figure}

Decreasing the metallicity in stellar cores results in a decrease in the internal opacity allowing energy to be more efficiently transported from the core. This allows a star of the same mass but lower metallicity to sustain a higher core temperature and thus produce more energy. The sensitivity of the $J_{f125w}-H_{f160w}$ colour to the metallicity is shown in the fourth panel of Figure \ref{fig:C} - decreasing the metallicity from $Z=0.02\to 0.004$  results in $\Delta(J_{f125w}-H_{f160w})\approx -0.08$. This is an important effect to take into account given observations of sub-solar metallicity UV producing stellar populations in galaxies at high-redshift (e.g. Pettini et al. 2000).

\subsection{Dust}\label{sec:dust}

By far the largest potential effect on the UV colour of star forming galaxies is the presence of large quantities of dust. The strong wavelength dependence of typical (e.g. Calzetti et al. 2000) reddening curves (where $k_{\lambda}$ is a monotonically decreasing function of $\lambda$) results in greater extinction in the UV (relative to the optical or near-IR) and the reddening of UV colours. The reddening of the UV colours then provides a potential diagnostic of the magnitude of the dust attenuation (Meurer et al. 1999). 

Applying the Calzetti et al. (2000) curve to the default scenario inferred SED produces a relationship between $E(B-V)_{\rm Calzetti}$ and the UV continuum ($C-D$) colour which is approximately linear, i.e. $E(B-V)\equiv d_{1}\times (C-D) +d_{2}$ (an expression similar to that suggested by Meurer et al. 1999). Differences between the relationships for different filter combinations arise because in each case the specific filter combination probes a unique range of the rest-frame UV continuum the values of $(d_{1}, d_{2})$ are $(1.22,-0.05)$, $(1.07,-0.04)$ and $(0.88,-0.04)$ for the $v$, $i$ and $z$ drops respectively.

Of course, a significant concern, is whether the Calzetti et al. (2000) curve is appropriate for high-redshift star forming galaxies, given that it was empirically derived from local starburst galaxies. At high-redshift (esp. $z>6.$) there has been insufficient time to produce significant amounts of dust through the winds of stellar atmospheres suggesting that any dust may be dominated by a supernovae component (Dunne et al. 2003) and may therefore have a different composition and thus reddening curve. However since the precise nature and UV-extinction properties of dust in starburst galaxies at this epoch is still poorly constrained, we use the Calzetti et al. (2000) dust curve as a first approximation.

\begin{figure*}
\centering
\includegraphics[width=35pc]{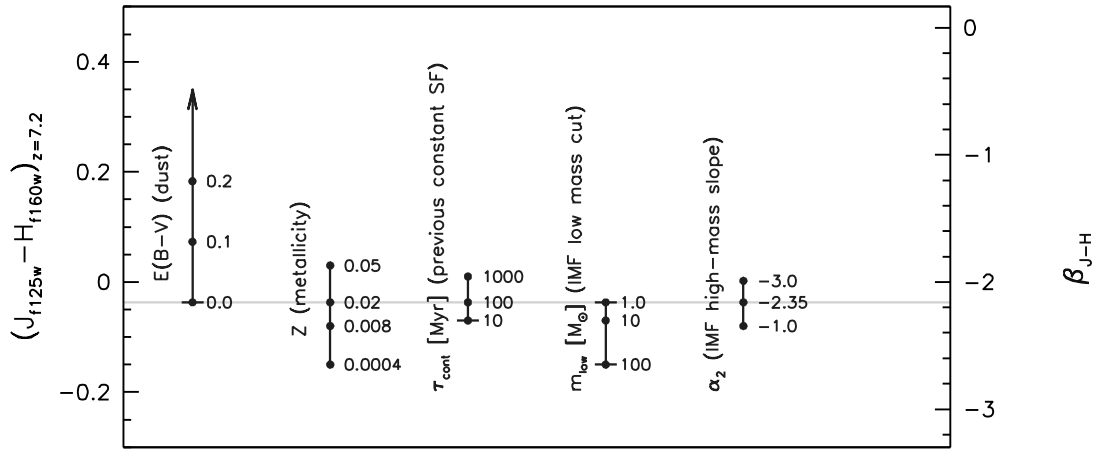}
\caption{A summary of factors affecting the UV continuum slope. Line caps denote the maximum deviation that are possible for each parameter. For example, Calzetti et al. (2000) dust is unable to produce $J_{f125w}-H_{f160w}<-0.037$ without changing any of the properties intrinsic to the stellar population from the {\em default} values. The horizontal line denotes the $J_{f125w}-H_{f160w}$ colour inferred from the {\sc Pegase.2} SPS model assuming the {\em default} scenario: $100\,{\rm Myr}$ previous continuous star formation, $Z=0.02$, a Salpeter IMF and no dust.}
\label{fig:sum}
\end{figure*}

Both the effect of dust (though only assuming the Calzetti et al. 2000 curve), and the other parameters described in the preceding section are summarised in Figure \ref{fig:sum}.

\subsubsection{The Interpretation of UV Colours}\label{sec:interp}

It is important to stress that these relationships (between the UV continuum colour and the colour excess) only hold for the {\em default} scenario defined earlier in this section (and of course are only relevant for the Calzetti et al. (2000) reddening curve). Any deviation in the underlying stellar properties (metallicity, SFH, IMF) will have a systematic effect on the dust properties derived from the rest-frame UV colour complicating the interpretation. Thus without additional information (which is scarce at high-redshift) any observed UV colour will be unlikely to have a unique interpretation. For example, while a $J_{f125w}-H_{f160w}=-0.037$ is consistent with the {\em default} scenario (i.e. no dust) it also consistent with non-zero attenuation together with lower stellar metallicity or a shorter duration of previous star formation. 

One exception to this is if extremely blue UV continuum colours are observed; while the UV continuum slope can be made arbitrarily red by adding additional dust, the same is not true in the opposite direction (i.e. the UV slope can not be made arbitrarily blue). For example, a $J_{f125w}-H_{f160w}<-0.2$ colour ($\beta<-2.9$) for $z$-drop selected galaxy is inconsistent (assuming the accuracy of the {\sc Pegase.2} model) with $Z>0.0004$ metallicity and any potential star formation history. A secure observation of such a colour may then potentially be an indication of an additional effect (see Bouwens et al. 2010a), perhaps due to stellar populations with ultra-low metallicity or a different IMF.

While it is not currently possible to independently and accurately constrain the metallicity and star formation histories of high-redshift galaxies this can be investigated in a statistical sense (i.e. for large samples of galaxies) using galaxy formation simulations. For example, galaxy formation simulations can be used can be used to determine the star formation and metal enrichment histories of galaxies exceeding some UV luminosity threshold. Using these properties combined with a population synthesis model can then be used to determine the contribution of the SFH and metallicity to the UV continuum colours. Of course such a technique is reliant on the accuracy of galaxy formation models, nevertheless it provides a first step in constraining the magnitude of these effects and thus ascertaining contribution from dust to UV continuum colour (Wilkins et al., {\em in prep}).

\section{UV Continuum Colours of Candidates}\label{sec:results}

Using data from the deep imaging fields discussed in Section 2 and the selection criteria laid out in Section 3 we identify $v$, $i$ and $z$ drop candidate high-redshift galaxies with $M_{1500}<-18.5$; the location of each galaxy in the $(A-B)$ - $(C-D)$ colour plane is shown in Figure \ref{fig:cc} together with the selection window and potential contaminants. 

Figure \ref{fig:mags} displays the $(C-D)$ colour for each candidate as a function of the absolute magnitude at $1500{\rm \AA}$. The absolute magnitude is derived from the magnitude in the $C$ filter corrected using the $(C-D)$ colour to $1500{\rm\AA}$. The rest frame location of the pivot point of the $C$ filter assuming the mean redshift are $1416{\rm \AA}$, $1521{\rm \AA}$ and $1524{\rm \AA}$ for the $v$, $i$ and $z$-drops respectively; thus the magnitude of this correction is typically very small. However, the uncertainty in the redshift introduces a systematic uncertainty on the absolute magnitude which is dependent on the $(C-D)$ colour (and its uncertainties).   

Throughout this section we employ the observed UV continuum colour (i.e. $(C-D)_{\rm AB}$), the $\beta$ parameter and the $E(B-V)$ inferred assuming the Calzetti et al. (2000) reddening curve and various assumptions regarding the stellar population properties (the {\em default} scenario: $100\,{\rm Myr}$ previous continuous star formation, solar $Z=0.02$ metallicity and a Salpeter IMF combined with the {\sc Pegase.2} population synthesis model).

\subsection{Very High Redshift Candidates}

There has recently been significant interest in the literature as to the security of many proposed high-redshift candidates. Given the limited data and the extremes to which the data is pushed this is unsurpring especially when using selection criteria aimed at maximising the number of candidates, and highlights the need for additional information (such as the detection of Lyman-$\alpha$ emission) to securely confirm the identity. 

For three reasons the very-high redshift objects considered in this work can be considered as more robust than Wilkins et al. (2011). Firstly, the criteria employed to unformly select galaxies irrespective of their UVC colour demands a redder break-colour ($A-B$) than employed previously, reducing the risk of contamination from the main galaxy sample due to photometric scatter (however it does increase slightly the risk of contamination from intrinsically red objects). Secondly we impose a stricter optical non-detection criteria compared to Wilkins et al. (2011) in the light of other work in this field and concerns over the robustness of faint, ambiguous sources, and finally we probe a slightly brighter range of galaxies (again reducing the contamination due to photometric scatter and spurious sources).

Indeed we see a decline in the number of candidates presented in Wilkins et al. (2011) to $9$,$4$ and $5$ in the HUDF, P12, and P34 fields respectively. This does not necessarily mean that the remaining candidates from Wilkins et al. (2011) are interlopers as many of the candidates are lost due to either being outside the new selection criteria (and thus redshift range) or fail to meet our brighter magnitude limit.

Comparing our candidates to other major studies: Bouwens et al. (2010c) and McLure et al. (2011), we find a high-level of agreement with $17$ of our $18$ included objects being included in one of the two secure candidate lists. Specifically in the HUDF field all $10$ of our objects are included in both Wilkins et al. (2011) and Bouwens et al. (2010c) and $9$ are included in McLure et al. (2011). In the P34 field $3$ of our $5$ candidates are included in primary listing of Bouwens et al. (2010c), while the additional two are listed as potential. $4$ of the $5$ objects are included in McLure et al. (2011), including the $2$ potential objects from Bouwens et al. (2010c). In the P12 field $2$ of our objects are included in the primary listing and the third is included as potential in Bouwens et al. (2010c) (McLure et al. 2011 does not analyse this field).

\subsection{Luminosity Dependence}\label{sec:res.mag}
\begin{table*}
\begin{tabular}{cccccccccccccc}
drop &  & $n$ & $\langle (C-D) \rangle$ & $\langle E(B-V) \rangle $ & $\langle \beta_{C-D} \rangle$ \\
\\
\hline
\hline


    & $-21.5<M_{1500}\le-20.5  $ & $  2  $ & $  0.07\,(<0.005)^{a}  $ & $  0.13\,(<0.005)  $ & $  -1.56\pm 0.02\,(0.02) $ \\
$v_{f606w}$    & $-20.5<M_{1500}\le-19.5  $ & $  11  $ & $  0.11\pm 0.05\,(0.18)  $ & $  0.18\pm 0.07\,(0.22)  $ & $  -1.32\pm 0.34\,(1.13) $ \\
    & $-19.5<M_{1500}\le-18.5  $ & $  23  $ & $  0.02\pm 0.04\,(0.19)  $ & $  0.07\pm 0.05\,(0.23)  $ & $  -1.89\pm 0.24\,(1.17) $ \\
\hline
    & $-21.5<M_{1500}\le-20.5  $ & $  4  $ & $  0.08\pm 0.02\,(0.03)  $ & $  0.13\pm 0.02\,(0.04)  $ & $  -1.58\pm 0.09\,(0.19) $ \\
$i_{f775w}$    & $-20.5<M_{1500}\le-19.5  $ & $  12  $ & $  -0.01\pm 0.05\,(0.17)  $ & $  0.03\pm 0.05\,(0.18)  $ & $  -2.04\pm 0.26\,(0.9) $ \\
    & $-19.5<M_{1500}\le-18.5  $ & $  36  $ & $  -0.05\pm 0.04\,(0.24)  $ & $  -0.02\pm 0.04\,(0.25)  $ & $  -2.29\pm 0.21\,(1.27) $ \\
\hline
    & $-21.5<M_{1500}\le-20.5  $ & $  1  $ & $  0.03^{b}  $ & $  0.07  $ & $  -1.88 $ \\
$z_{f850lp}$     & $-20.5<M_{1500}\le-19.5  $ & $  7  $ & $  -0.14\pm 0.05\,(0.13)  $ & $  -0.08\pm 0.04\,(0.11)  $ & $  -2.6\pm 0.21\,(0.55) $ \\
    & $-19.5<M_{1500}\le-18.5  $ & $  10  $ & $  -0.06\pm 0.06\,(0.2)  $ & $  -0.01\pm 0.06\,(0.17)  $ & $  -2.26\pm 0.27\,(0.85) $ \\
\hline

\end{tabular}
\caption{The number of candidates $n$, mean, standard error on the mean $\sigma/\sqrt{n}$ and standard deviation (in brackets) of the $(C-D)$ colour, the inferred value $E(B-V)_{\rm Calzetti}$ and $\beta$ parameters for the three magnitude bins in each drop-out sample. $^{a}$ the small number of objects which with similar colors results in a small $(<0.005)$ standard deviation. $^{b}$ only one object in the bin.}
\label{tab:mags}
\end{table*}

\begin{figure}
\centering
\includegraphics[width=20pc]{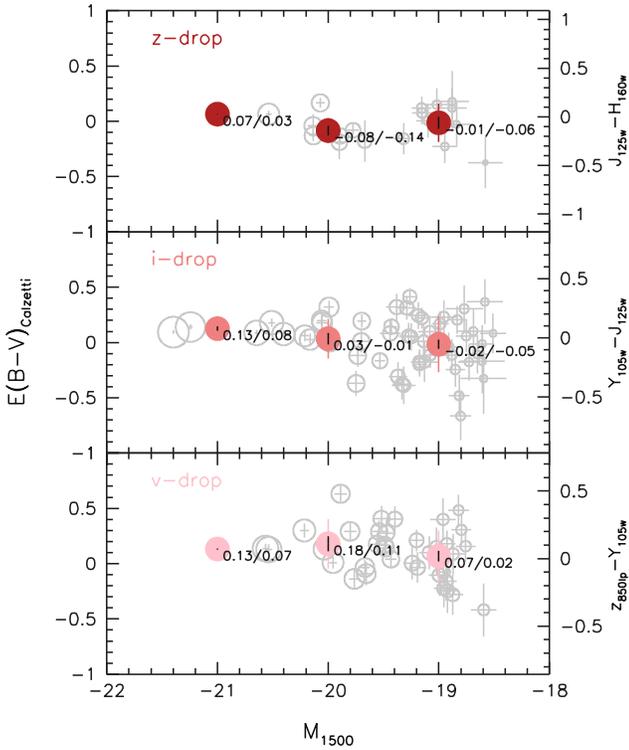}
\caption{The distribution of $(C-D)$ colours and the inferred $E(B-V)$ values as a function of absolute magnitude ($M_{1500}$) for candidates with $M_{1500}<-18.5$. The three panels are for the three dropout samples with $z$, $i$ and $v$ the top, middle and bottom respectively. The large filled points denote the mean $(C-D)$ colour for three bins and labels indicate the $E(B-V)_{\rm Calzetti}$ and $(C-D)$ colour on the left and right respectively. The large error bars denote the standard deviation $\sigma$ of the objects around the mean while the small black bars denote the standard error on the mean ($\sigma/\sqrt{n}$).}
\label{fig:mags}
\end{figure}

The absolute magnitude (or luminosity dependence) of the UV colours can be most easily explored by binning the galaxies by their absolute magnitude. In Figure \ref{fig:mags} the mean $(C-D)$ colour (and $E(B-V)_{Calzetti}$ value) for 3 absolute magnitude bins ($-21.5 < M_{1500} \le -20.5$, $-20.5 < M_{1500} \le -19.5$ and $-19.5 < M_{1500} \le -18.5$) for each of the drop-out sample is shown. The values of the mean $(C-D)$ colour, $E(B-V)_{Calzetti}$ and $\beta_{C-D}$ together with the standard deviation are also presented in Table \ref{tab:mags}. 

For both the $v$ and $i$-drops we find that the mean $(C-D)$ colour becomes is redder in the high and intermediate luminosity bins compared to the $-19.5 < M_{1500} \le -18.5$ bin (while the brightest $v$-drop bin is bluer than the intermediate bin it only contains $2$ objects and is thus susceptible to large uncertainties). This behaviour is consistent with the general trend found by other studies (e.g. Bouwens et al 2009,2010a) at these and lower redshifts. For the high redshift $z$-drops the trend is less clear, as the intermediate magnitude bin has a mean $(C-D)$ colour bluer than both the high and low-luminosity bins. Of course, the analysis of the $z$-drops is affected by the small number of galaxies available. If instead 2 bins are used ($-21.5 < M_{1500} \le -20.0$ and $-20.0 < M_{1500} \le -18.5$) there is the indication of a similar trend found for the $v$ and $i$ drops. It is important to remember that the faintest $z$-drop bin is potentially biased against red $(C-D)$ colours; the implication of this is discussed in more detail in the following section where the observed distribution of $(C-D)$ colours is analysed.

\subsubsection{Redshift Evolution}\label{sec:res.z}

\begin{figure}
\centering
\includegraphics[width=20pc]{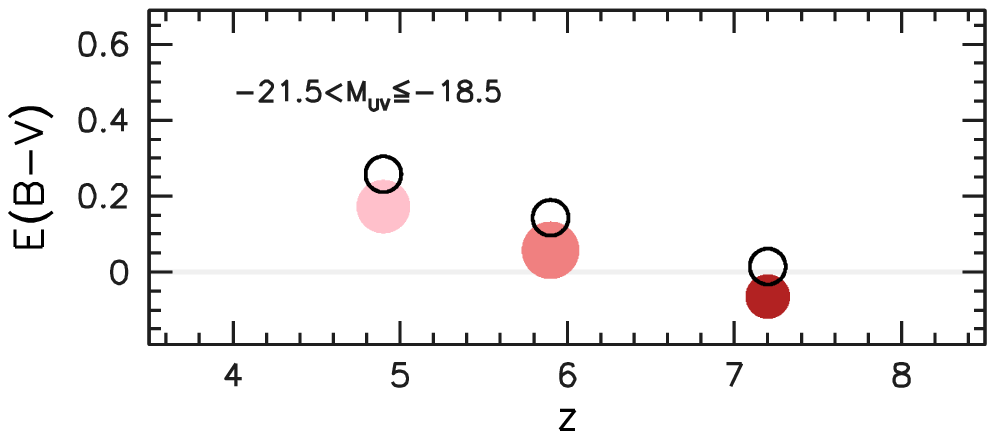}
\includegraphics[width=20pc]{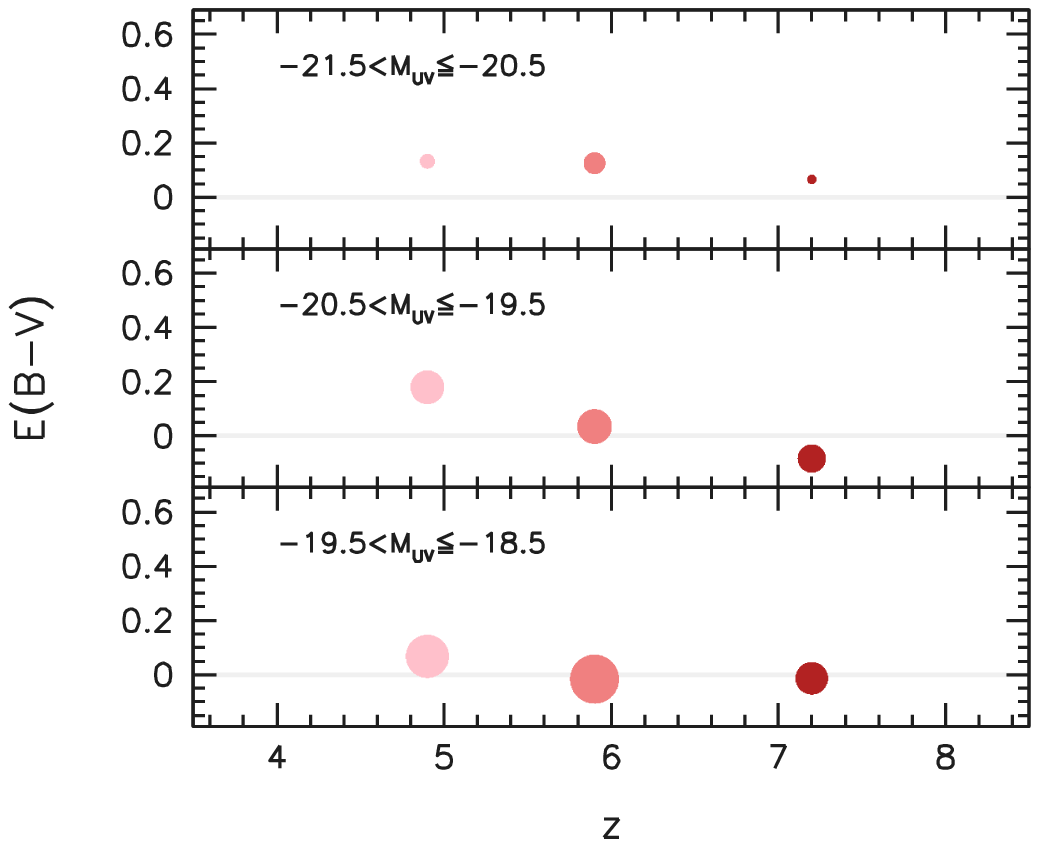}
\caption{The redshift evolution of the mean value of $E(B-V)$ and $\beta$ for candidates with $-21.5 < M_{1500}\le -18.5$ (top panel) and for each of the three Absolute magnitude bins considered in Section \ref{sec:res.mag}. The size of the points in each panel are proportional to the $\log_{10}$ of the number of objects. The open circles in the top panel denote the values of $E(B-V)$ inferred assuming the metallicity evolution predicted from galaxy formation simulations (see \S\ref{sec:interp.Zevo}).}
\label{fig:z}
\end{figure}

The top panel of Figure \ref{fig:z} shows the mean $E(B-V)$ for galaxies with $-21.5 < M_{1500} \le -18.5$ in each drop sample highlighting the redshift evolution of the rest-frame UV colours. A significant redshift evolution is found whereby the mean $(C-D)$ colour (and thus the mean value of $E(B-V)$ inferred assuming the {\em default scenario}) becomes redder at lower redshift. However, this is potentially biased as the relative effective volumes for the low and high-luminosity varies for each dropout sample (this is seen in Figure \ref{fig:vol}). Specifically the lower-redshift $v$ and $i$ drops have a different {\em theoretical} luminosity distribution from the $z$-drops. 

To avoid this bias, and to investigate the luminosity dependence of the redshift evolution, we can split each sample into the three absolute magnitude bins previously considered; while this allows us to tackle the luminosity distribution bias, the uncertainties increase due to the small number of galaxies in each bin. In the intermediate magnitude bin ($-20.5 < M_{1500} \le -19.5$) there is again evidence for redshift evolution such that at lower redshift galaxies typically have redder UV continuum slopes. In the faintest bin both the $i$ and $z$-drop selected samples have colours $\simeq -0.05$ while the lower redshift $v$-drops are slightly redder. The brightest bin again shows evidence for some variation but is severely limited by the small numbers of galaxies.

\subsection{Intrinsic Distribution of UV colours}\label{sec:res.int}

\begin{figure*}
\centering
\includegraphics[width=40pc]{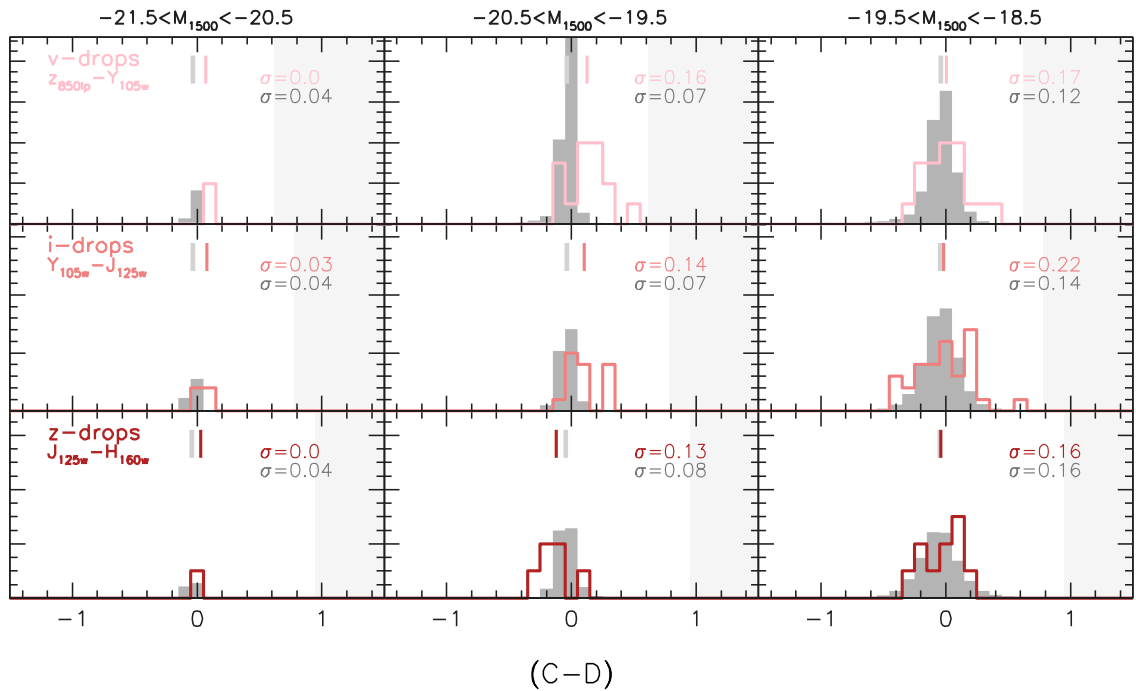}
\caption{Observed (lines) and simulated (solid) histograms of galaxies binned by their UV continuum colour ($(C-D)$, see Table \ref{tab:params} for the corresponding $C$ and $D$ filters for each drop sample) for each drop sample (progressing vertically) and absolute magnitude bin (horizontally). The simulated histogram is the result of inserting synthetic galaxies into the original science frame before candidate selection. These galaxies have synthetic spectra assembled using the {\sc Pegase.2} population synthesis model (assuming $100 Myr$ of previous star formation, Salpeter IMF, solar metallicity and no dust -  the default scenario outline in Section 4). The deviation from a single observed value is then due entirely to photometric scatter and not a distribution of the intrinsic properties. The small vertical lines at the top of each panel denote the mean colours for both the simulated (grey) and observations distributions. The standard deviations of both the observed (upper label) and simulated distributions (lower label) are also shown in each figure. The shaded region on the right of each panel denotes $E(B-V)_{\rm Calzetti}>0.8$ which we do not probe.}
\label{fig:dist}
\end{figure*}

The distribution of rest-frame UV colours for each magnitude bin and dropout sample are shown in Figure \ref{fig:dist}. With the exception of the most luminous bin ($-21.5<M_{1500}<-20.5$), where there are too few objects ($n \sim 1-2$) to accurately quantify the scatter, we find that standard deviation of the observed colour distributions to be $\sigma\sim 0.13-0.22$. It is interesting to consider how much of this scatter is caused by an intrinsic distribution in the properties of the galaxies (including dust) and not just due to photometric noise. 

To asses this a series of simulations are performed where we introduce {\em synthetic} sources into the original images, similar to the technique used in Section \ref{sec:effvol}. In this case sources are inserted with a single rest-frame UV continuum colour corresponding to a synthetic spectrum assuming the {\em default} parameters defined in Section \ref{sec:uv}. We then run the same source identification/photometry/selection as used to identify our candidates.

These simulated {\em single-model} distributions are shown in Figure \ref{fig:dist} as solid histograms for each dropout sample and each magnitude bin. The standard deviation of the recovered distribution increases at fainter magnitudes (and higher redshift for the same absolute magnitude) reflecting the expected increase in photometric noise.

For the faintest magnitude bin ($-19.5<M_{1500\AA}<-18.5$) we find that the standard deviation of the observed distribution for the $v$ and $i$-drops is slightly larger than that predicted solely from photometric noise, and as noted in Section \ref{sec:res.mag} they are slightly redder than the default scenario. This suggests that for these galaxies there is some intrinsic distribution of star formation histories, metallicities or dust content. For the faintest $z$-drop bin both the mean and standard deviation of the observed distribution are consistent with the simulated distribution; i.e. the observed distribution is consistent with a single intrinsic colour. The inferred narrow range of intrinsic properties suggests that the effects of the colour bias that was found to affect the faintest $z$-drop bin should not have a significant impact on the observed mean colour.

For the intermediate magnitude bin ($-20.5<M_{1500\AA}<-19.5$) the disparity between the single-model simulation and the observed distribution increases; the standard deviation of the distribution of the observed galaxies is larger than that predicted from the simulations (i.e. $\sigma_{\rm obs}>\sigma_{\rm sim}$). As noted in Section \ref{sec:res.mag} the mean colours of the $v$ and $i$-drops in these bins are also redder than that predicted assuming the default parameters. Again, this suggests that these galaxies have a distribution of at least one of the star formation history, IMF, metallicity or dust content.

\subsection{The Physical Interpretation of UV Continuum Colours}

As illustrated in Section \ref{sec:interp} the interpretation of the UV colours of star forming galaxies is ambiguous as the continuum is affected not only by dust but also by the recent star formation history and the metallicity of the UV producing population. However, changes to the metallicity and star formation history cannot produce arbitrarily blue intrinsic UV continuum slopes, and it is difficult to produce extremely red {\em intrinsic} slopes while still recording significant UV luminosity. 

By assuming an alternative scenario with extremely low ($Z=0.0004$), though {\em non}-zero, metallicity, a short duration of previous star formation ($10\,{\rm Myr}$) and a Salpeter IMF, we can place upper limits on the amount of dust attenuation. For all (i.e. $M_{1500}<-18.5$) the $v$, $i$ and $z$ drops the upper-limits, assuming this low metallicity, on the mean value of $E(B-V)_{\rm Calzetti}$ are $0.32$, $0.20$ and $0.06$ respectively (c.f. $E(B-V)_{\rm Calzetti}$ assuming $Z=0.02$: $0.17$, $0.06$ and $-0.06$). 

\subsubsection{Metallicity Evolution}\label{sec:interp.Zevo}

Over the period spanned by $z=7.7 \to 4.7$ the metallicity of the UV producing population is expected to increase as core collapse supernovae inject enriched material into the interstellar medium. As we saw in Section \ref{sec:interp.dust} lowering the metallicity shifts the intrinsic colour of the stellar population blueward. Thus including the effect of an {\em increasing} metallicity will then act to slightly flatten the inferred trend of $E(B-V)$. To explore this in more detail we take advantage of galaxy formation simulations (e.g. Dayal et al. 2009 and Dayal, Ferrara \& Saro 2010) that predict the evolution of the metallicity of the UV producing population relevant for our sample. For galaxies selected as $v$, $i$ and $z$-drops, with $M_{1500}<-18.5$ average metallicities of $Z/Z_{\odot}=0.22$, $0.18$ and $0.12$ are found respectively (P. Dayal, {\em Priv. Comm.}). Taking into account these metallicities in the conversion of the observed $(C-D)$ colour to the colour excess assuming Calzetti et al. (2000) we now find  $E(B-V)_{\rm Calzetti}\simeq 0.25$, $0.14$ and $0.01$ for the $v$, $i$ and $z$ respectively. These values of $E(B-V)_{\rm Calzetti}$ are shown alongside those inferred assuming the {\em default} scenario in the top panel of Figure \ref{fig:z} to highlight the increase.

\subsection{Comparison with Previous Studies}

\subsubsection{Bouwens et al. 2010a}

The analysis of Bouwens et al. (2010a), based on {\em HST} WFC3 observations of the HUDF, finds a significant luminosity dependence and, of greater interest, extremely blue ($\beta\simeq -3$, $(J_{f125w}-H_{f160w})\simeq -0.2$) UV colours of the least luminous ($M_{1500}\simeq -19.0$) $z$-drop sources. These are interpreted as evidence of an extremely low metallicity population. While we can not exclude a luminosity dependence of the UV continuum colour for our $z$-drop sources we do not find significant numbers of extremely blue sources at these faint luminosities. While we observe some galaxies with $\beta\simeq -3.0$, the results of our simulations (Section \ref{sec:res.int}) suggest these are consistent with that expected due to photometric scatter. The principal difference between our analysis and that of Bouwens et al. (2010) is the choice of selection criteria; while ours is designed to minimise the UV colour bias that of Bouwens et al. (2010) is optimised to identify a maximum number of robust $z\simeq 7$ star forming galaxy candidates.

\subsubsection{Dunlop et al. 2011}

A similar analysis, based on {\em HST} WFC3 observations, of the UV colours of galaxies has also recently been undertaken by Dunlop et al. (2011) using a photometric redshift based selection (see McLure et al. 2011) instead of the colour-colour-colour selection utilised in this study. Dunlop et al. (2011) finds UV colours over the redshift range $z=5-7$ consistent with $\beta\simeq -2.$ (i.e. $(C-D)\simeq 0.0$). While these results are consistent with our analysis for the $i$ and $z$ drops we find somewhat redder UV continuum slopes for the lower-redshift $v$ drops ($\langle \beta \rangle |_{v}\simeq -1.7\pm 0.3$), though note our results are consistent within the mutual $1\sigma$ uncertainty intervals.

\subsubsection{Bouwens et al. 2009}

The study of Bouwens et al. (2009) measures the UV continuum slope over $z\simeq 2\to 6$ based on {\em HST} ACS and NICMOS observations in the vicinity of GOODS-South and North. While both our measurements, and those of Bouwen et al. (2009) are consistent at $z\simeq 6$, we find somewhat redder UV continuum slopes at $z\simeq 5$, especially at lower luminosities where we find $\beta\simeq -1.89\pm 0.24$ at $M_{1500}\simeq -19.0$ while Bouwens et al. (2009) finds $\beta\simeq -2.64$. The reasons for this difference are unclear but again are likely be due to the bias against redder UV colours arising from the difference in the choice of selection criteria. In addition however there are number of other possibilities; firstly the Bouwens et al. (2009) study is largely based on different samples, suggesting the possibility of cosmic variance. Secondly the Bouwens et al. (2009) analysis is based on NICMOS observations (as opposed to WFC3), raising the possibility of errors due to uncertainties on the respective zeropoints. The use of different filters also means we are probing different ranges of the UV continuum and finally there remains the possibility that one, or both, studies suffer from some degree of contamination potentially biasing the mean UV continuum colours in one direction.

\section{Conclusions}

In this work we have used {\em HST} WFC3 imaging of the Hubble Ultra Deep Field (HUDF) and two associated parallel deep fields to measure the rest-frame UV continuum properties of high redshift galaxies ($z>4.5$). 

This sample of galaxies was selected using a Lyman-break technique and was split into three redshift intervals corresponding to objects {\em dropping out} in the $v_{f606w}$, $i_{f775w}$ and $z_{f850lp}$-filters. This selection criteria employed is tailored to {\em uniformly} select galaxies for a large range of intrinsic rest-frame UV continuum colours, thereby intending to mitigate any non-explicit biases.   

For both the $v$ and $i$-drop selected samples we find evidence, as also reported by other studies (e.g. Bouwens et al. 2009), for a luminosity dependence in the mean rest-frame UV continuum colours. Specifically, more luminous objects are found to have redder colours. For the $z$-drop selected sample the trend is less clear due to the small numbers of galaxies in each luminosity bin. 

In addition we see evidence for evolution in the rest-frame UV colours $z\simeq 4.7\to 7.7$ such that galaxies at lower-redshift (over the same luminosity range) are typically redder. While this is in general agreement with other studies we observe a stronger degree of evolution.

By comparing the observed distribution of colours with the results of photometric scatter simulations we also investigate the intrinsic distribution of UV colours. We find the observed distribution of the least-luminous $z$-drop selected sample is consistent with a single intrinsic colour corresponding to a dust-free star forming population with $Z\simeq 0.02$ and a previous constant star formation history of $100\,{\rm Myr}$. This contradicts the findings of Bouwens et al. (2010) who report extremely blue UV ($\beta\simeq -3.0$) UV continuum slopes. A potential reason for this discrepancy is use of a selection critieria biased against red UV continuum slopes, and assertion supported by Dunlop et al. (2011). At lower redshifts, and increasing with luminosity, the observed distribution of colours widens such that it is no longer consistent solely with photometric scatter ($\sigma_{\rm obs}>\sigma_{\rm sim}$) suggesting that galaxies at these redshifts have a range of intrinsic rest-frame UV continuum colours. 

The common interpretation of variation of the rest-frame UV continuum colours is that they are due to differences in the amount of dust attenuation. While this is likely to be the predominant explanation the rest-frame UV continuum colours of stellar populations are also affected by the star formation history and metallicity (and initial mass function), parameters which are difficult to independently constrain. Ascertaining the contribution of these effects is challenging from observations, especially at high-redshift. Galaxy formation models combined with population synthesis models may be capable of providing estimates for the extent of these effects allowing the determination of the remaining colour deviation due to the effects of dust.

{\em HST} WFC3 NIR programmes in progress (e.g. the CANDELS MCT project), together with ground based NIR surveys (e.g. UltraVISTA) will provide further constraints on the rest-frame UV properties of the luminous high-redshift galaxy population by increasing the sample size and covering a wider range in luminosities. However due to the limited spectral coverage available at wavelengths greater than $1.8\mu m$ (i.e. beyond the $H_{f160w}$ filter) studies of the UV properties of $z>8$ objects will {\em likely} be limited until the deployment of the James Webb Space Telescope ({\em JWST}). {\em JWST} will provide not only the increased wavelength coverage to probe the $z>8$ Universe but also the survey efficiency to discover and precisely characterise large numbers of galaxies at lower redshifts ($z>3$).

\subsection*{Acknowledgements}
We would like to thank St\'{e}phane Charlot, Camilla Pacifici, Jacopo Chevallard and Richard Ellis for useful comments in the preparation of this manuscript. SMW acknowledges the support of STFC. Based on observations made with the NASA/ESA Hubble Space Telescope,
obtained from the Data Archive at the Space Telescope Science Institute, which is operated by the Association
of Universities for Research in Astronomy, Inc., under NASA contract
NAS 5-26555. These observations are associated with programme \#GO-11563.

\bsp

\end{document}